\documentclass[sigconf,final,nonacm]{acmart}

\usepackage{amsmath,amsfonts}
\usepackage{algorithmic}
\usepackage{graphicx}
\usepackage{textcomp}
\usepackage{subcaption}
\usepackage{tabularx}
\usepackage{xcolor}
\usepackage{color,soul}
\usepackage{xspace}
\usepackage[inline]{enumitem} 
\usepackage{booktabs}
\usepackage{multirow}
\usepackage{graphicx}
\usepackage{hyperref}
\graphicspath{{img/}}
\usepackage{enumitem}
\usepackage{pifont}
\usepackage{listings}
\usepackage{booktabs}
\usepackage{tcolorbox}
\usepackage[multiple]{footmisc}

\newcommand{\NAME}[0]{GitRanking\xspace}

\newtcolorbox[auto counter]{mybox}[1][]{%
    title=Takeaway~\thetcbcounter, arc=1pt,outer arc=1pt, #1}
\AtBeginDocument{%
  \providecommand\BibTeX{{%
    \normalfont B\kern-0.5em{\scshape i\kern-0.25em b}\kern-0.8em\TeX}}}

\lstdefinelanguage{m3}{
	basicstyle=\ttfamily\scriptsize,
	keywordstyle=\bfseries,
	keywords={m3,declarations,methodInvocation},
	literate={<-}{$\leftarrow$}{1},
	tabsize=2,
	alsoletter={-}
}

\newtcolorbox{shadedbox}{
	drop shadow southeast,
	breakable,
	enhanced jigsaw,
	colback=white,
	boxrule=0.80pt
}

\usepackage{xspace}
\newcommand*{\ie}{i.e.,\@\xspace}
\newcommand*{\eg}{e.g.,\@\xspace}

\makeatletter
\newcommand*{\etc}{%
	\@ifnextchar{.}%
	{etc}%
	{etc.\@\xspace}%
}
\makeatother

\definecolor{darkgray}{gray}{0.78}
\definecolor{lightgray}{gray}{0.85}
\definecolor{verylightgray}{gray}{0.95}
\begin{document}
\title{\NAME: A Ranking of GitHub Topics for Software Classification using Active Sampling}
\author{Cezar Sas}
\email{c.a.sas@rug.nl}
\orcid{0000-0002-3018-0140}
\author{Andrea Capiluppi}
\orcid{0000-0001-9469-6050}
\email{a.capiluppi@rug.nl}
\affiliation{%
  \institution{University of Groningen - Bernulli Institute}
  \streetaddress{Nijenborgh 9}
  \city{Groningen}
%  \state{Groningen}
  \country{Netherlands}
  \postcode{9747 AG}
}

\author{Claudio Di Sipio}
\email{claudio.disipio@graduate.univaq.it}
\orcid{0000-0001-9872-9542}
\author{Juri Di Rocco}
\orcid{0000-0002-7909-3902}
\email{juri.dirocco@univaq.it}
\author{Davide Di Ruscio}
\orcid{0000-0002-5077-6793}
\email{davide.diruscio@univaq.it}
\affiliation{%
  \institution{University of L'Aquila}
  \streetaddress{Via Vetoio}
  \city{L’Aquila}
  \country{Italy}
  \postcode{67100}
}

\renewcommand{\shortauthors}{Sas C., et al.}

\begin{abstract}
 
GitHub is the world's largest host of source code, with more than 150M repositories. However, most of these repositories are not labeled or inadequately so, making it harder for users to find relevant projects.
There have been various proposals for software application domain classification over the past years. However, these approaches lack a well-defined taxonomy that is hierarchical, grounded in a knowledge base, and free of irrelevant terms.

This work proposes \NAME, a framework for creating a classification ranked into discrete levels based on how general or specific their meaning is. We collected 121K topics from GitHub and considered 60\% of the most frequent ones for the ranking. \NAME 1) uses \textit{active sampling} to ensure a minimal number of required annotations; and 2) links each topic to Wikidata, reducing ambiguities and improving the reusability of the taxonomy. 

Our results show that developers, when annotating their projects, avoid using terms with a high degree of specificity. This makes the finding and discovery of their projects more challenging for other users. Furthermore, we show that \NAME can effectively rank terms according to their general or specific meaning. This ranking would be an essential asset for developers to build upon, allowing them to complement their annotations with more precise topics.

Finally, we show that \NAME is a dynamically extensible method: it can currently accept further terms to be ranked with a minimum number of annotations (\textasciitilde 15). This paper is the first collective attempt to build a ground-up taxonomy of software domains.

\end{abstract}





\maketitle

\section{Introduction}
\label{sec:intro}
GitHub is the world’s largest host of source code, with more than 150M repositories in 2021; moreover, the number of repositories increased by 60M+ in the previous year\footnote{\href{https://octoverse.github.com/}{https://octoverse.github.com/}}. However, these repositories are not easy to find: while GitHub allows developers to annotate their projects manually and other users to search software via \textit{Topics}\footnote{\href{https://github.com/topics}{https://github.com/topics}}, not all projects are making use of it, or use it inefficiently, by just annotating with one or two topics. Additionally, developers are free to annotate a project with any string they want: this inevitably generates a very large number of specific, unrepresentative labels. 

Various works in the literature have attempted to automatically classify the domains of software applications, many proposing their datasets with custom taxonomies~\cite{sharma2017cataloging, vargas2015automatic, zhang2019HiGitClass}, and more recently, using a subset of GitHub Topics~\cite{sipio2020naive, 9590294}.
However, these resources suffer from various recurring problems, the antipatterns of software classifications~\cite{sas2022antipatterns}. First, no current taxonomy explicitly defines a hierarchical relation among their labels, making it problematic when dealing with their single label annotation and `\texttt{IS-A}' relationships among labels (\textit{mixed granularity} issue). A second issue is the mix of different taxonomies in the same categorization (\textit{mixed taxonomies} issue), for example, when labels from application domains (\eg `\textit{Security}') are present as well as programming languages (\eg `\textit{Python}'). This is an issue when performing single task classification as opposite to multi-task~\cite{caruana1997mtl}, which would not be possible given the lack of separation between the taxonomies. Furthermore, these categorizations are not complete, as they do not cover the entire spectrum of software categories (\eg having `\textit{Compiler}' but not `\textit{Interpreter}'), making it easier for a model to distinguish some classes. The incompleteness is also aggravated by the fact that no work is grounded to a knowledge base (KB): this is highly problematic because it does not resolve the ambiguity of an arbitrarily defined categorization (\ie top-down), reducing its usability and the possibility to add new terms to it (\textit{ambiguity} issue). All the issues above make these categorizations less valuable to be used in a real-world scenario.

As an alternative to the pre-defined taxonomies presented by previous works in software application domain classification, in the Natural Language Processing field, there are works focusing on taxonomy construction from data~\cite{yu-etal-2020-hearst, shang-etal-2020-taxonomy}. While some solutions focus on creating a taxonomy for the Computer Science domain using papers from the bibliography service DBLP~\cite{zhang2018taxogen, shang2020nettaxo}, our attempt at reproducing their results failed. Furthermore, current solutions are not deterministic, requiring multiple runs and a lucky seed to get a good starting point for annotators to work on. Also, these and also other solutions require a large amount of data to create the taxonomy~\cite{wang-etal-2017-short}; however, this is not always available for GitHub Topics.

In this paper, to solve the issues outlined above, we present \NAME, a framework for ranking software application domains. In contrast to previous work, we defined a pipeline for selecting the topics, using 121K GitHub Topics as our initial seed. \NAME uses an active sampling method combined with a Bayesian inference algorithm to create the ranking of the topics. Furthermore, to reduce the intrinsic ambiguity of natural language, and make the taxonomy more usable, each term is linked to its Wikidata entity. One key feature of \NAME is the ability to easily expand the ranked taxonomy with a minimal amount of examples (\textasciitilde 15 for each new topic added). 

Furthermore, \NAME allowed us to extract insights regarding the usage of GitHub Topics by the practitioners, and answer the following research question: 
\smallskip
\begin{quote}
    \textit{\textbf{RQ}: Are the topics used to annotate GitHub projects evenly distributed in the levels of a taxonomy?}
\end{quote}
\smallskip

The ambition of these results is to help developers in better annotating their projects, and to make them easier to find and more discoverable. This improved discoverability will also help other developers, as it will be easier and faster to find the best library for a specific task improving the reusability of software. 

%
%\footnote{\href{https://github.com/SasCezar/GitHubClassificationDataset}{https://github.com/SasCezar/GitHubClassificationDataset}}
%While there are various solution for taxonomy induction and construction, these solution require large amounts of data to learn . Furthermore, the tools released are not deterministic, meaning that for each run they will return different taxonomies, depending heavily on the seed. 
%

In summary, our contributions are:

\begin{itemize}
    \item[--] An online framework for better-creating software categorization, and expand them; 
    \item[--] A list of 301 application domains extracted from GitHub Topic and disambiguated by linking them to Wikidata;
    \item[--] A ranking of the 301 topics into discrete levels based on their meaning,
    \item[--] Using our ranking, answer \textbf{RQ}.
%    \item Answer \textbf{RQ1}.
\end{itemize}

% Additionally, the collected data was used to assess this research question:

We made our code\footnote{\href{https://anonymous.4open.science/r/GitRanking/}{https://anonymous.4open.science/r/GitRanking/}} and data\footnote{\href{https://zenodo.org/record/5879573}{https://zenodo.org/record/5879573}} available.

\smallskip
This paper is articulated as follows: in Section~\ref{sec:_relworks} we analyze the past works in terms of existing taxonomies, and the approaches that were used to extract one from data. In Section~\ref{sec:_bkground} we present the ASAP (\textit{Active SAmpling for Pairwise comparisons}) algorithm, which we used for reducing the annotations required for the ranking, and the TrueSkill ranking system which creates a ranking of the annotated topics. Section~\ref{sec:_approach} describes the pipeline of \NAME, and its activities. Section~\ref{sec:_results} presents the results of the work performed by the annotators, and the TrueSkill output: we discuss these findings in Section~\ref{sec_discussion}. We analyze the threats to validity that we encountered in Section~\ref{sec:_threats}. We present the conclusion and future works in Section~\ref{sec:_conclusion}.

\section{Related Work}
\label{sec:_relworks}

In this section we present the relevant related work. We address works regarding the software classification problem, in particular application domain classification, and works focusing on taxonomy construction or induction.

\subsection{Software Classification Taxonomies}

There have been various attempts in the literature focusing on software classification, from application domains~\cite{vasquez2014api, sipio2020naive}, to bugs~\cite{moustafa2018software}, and vulnerabilities~\cite{sabetta2018security}. In this paper, we focus on works performing software application domains classification.  

One of the initial works on software classification is MUDABlue~\cite{Kawaguchi2006MUDABlue}. They propose a dataset of 41 projects written in C and divided into six categories. They also present a model based on information retrieval techniques, specifically Latent Semantic Analysis (LSA),  to classify software based on their source code identifiers.

Tian et al. proposed LACT~\cite{tian2009lact}. As for MUDABlue, the authors propose both a new dataset and a new approach to classification. The dataset consists of 43 examples divided into 6 SourceForge categories. The list of projects is available in their paper. Their classification model combines Latent Dirichlet Allocation (LDA), a generative probabilistic model that retrieves topics from textual datasets, and heuristics. They use the identifiers and comments in the source code as input to their model. 

Again, in~\cite{vasquez2014api}, the authors propose a new dataset using SourceForge as seed. The dataset consists of words extracted from API packages, classes, and methods names using naming conventions. However, the dataset containing 3,286 Java projects annotated into 22 categories is no longer available. Using the example in~\cite{Ugurel2002classification}, the authors use information gain to select the best attributes as input to different machine learning methods.

LeClair et al.~\cite{leclair2018neural} propose a dataset of C/C++ projects from the Debian package repository. The dataset consists of 9,804 software projects divided into 75 categories: many of these categories have only a few examples, and 19 are the same categories with different surface forms, more specifically `contrib/X', where X is a category present in the list. For the classification, they used a neural network approach. The authors use the project name, function name, and function content as input to a C-LSTM~\cite{zhou2015clstm}, a combined convolutional and recurrent neural networks model.

In~\cite{vargas2015automatic} authors proposed an approach to generate tag clouds starting from bytecodes, external dependencies of projects, and information extracted from Stack Overflow. Unfortunately, their dataset is not available.  

Sharma et al.~\cite{sharma2017cataloging} release a list of 10,000 examples annotated by their model into 22 categories, evaluated using 400 manually annotated projects. It is interesting to notice that half of the projects eventually end up in the `{Other}' category, which means that they are not helpful when training a new model. They used a combined solution of topic modeling and genetic algorithms called LDA-GA for the classification~\cite{panichella2013ldaga}. The authors apply LDA topic modeling on the \texttt{README} files and optimize the genetic algorithms' hyper-parameters. While LDA is an unsupervised solution, humans are needed to annotate the topics from the identified keywords. 

In ClassifyHub~\cite{soll2017classifyhub}, the authors use the InformatiCup 2017\footnote{\href{https://github.com/informatiCup/informatiCup2017}{https://github.com/informatiCup/informatiCup2017}} dataset, which contains 221 projects unevenly divided into seven categories. For the classification, they propose an ensemble of 8 na\"{i}ve classifiers, each using different features (\eg file extensions, \texttt{README}, GitHub metadata and more). 

In \cite{zhang2019HiGitClass}, the authors release two datasets spanning two domains: an artificial intelligence taxonomy with 1,600 examples and a bioinformatics one with 876 projects. The datasets have been annotated according to a hierarchical classification that is given as an input with keywords for each leaf node. Furthermore, they propose HiGitClass, an approach for modeling the co-occurrence of multimodal signals in a repository (\eg user, repository name, \texttt{README}, and more) to perform the classification.  %These taxonomies are contain of problem domain categories.

Focusing on unsupervised approaches, we find CLAN~\cite{mcmillan2012clan}, which provides a way to detect similar apps based on the idea that similar apps share some semantic anchors. They also propose a dataset (not available anymore) in previous work. Given a set of applications, the authors create two terms-document matrices: the structural information using the package and API calls, and the other for textual information using the class and API calls. Both matrices are reduced using LSA, then the similarity across all applications is computed. Lastly, the authors combine the similarities from the packages and classes by summing the entries. 
In \cite{linares2016clandroid}, the authors propose CLANdroid, a CLAN adaptation to the Android apps domain, and evaluate the solution on 14,450 Android apps. Unfortunately, their dataset is not available.

Another unsupervised approach was adopted by LASCAD~\cite{altarawy2018lascad}. However, unlike other unsupervised methods, the authors proposed an annotated dataset consisting of 103 projects divided into six categories (from GitHub Collections) with 16 programming languages (although many languages have only 1 example) and an unlabeled one which is not available. Their approach uses a language-agnostic classification and similarity tool. As in LACT, the authors used LDA over the source code and further applied hierarchical clustering with cosine similarity on the output topic terms matrix of LDA to merge similar topics. 

% Taking a more Natural Language Processing (NLP) inspired approach, based on the distributional hypothesis: `\textit{A word is characterized by the company it keeps}'~\cite{firth1957studies}, \cite{theeten2019import2vec} proposed a neural network solution to create dense representation (\ie embeddings) of libraries. The authors used the co-occurrences of import statements of libraries to learn a semantic space where libraries that appear in the same context are close (similar) in the vector space. The authors do not perform classification; therefore, their dataset is not annotated; however, the learned representation (available) can be used to compute similarity and train a classification model. 

More recent works focus on using GitHub as the source of their classification. In Di Sipio et al.~\cite{sipio2020naive}, the authors released a dataset for multi-label classification annotated with 120 popular topics from GitHub. The dataset contains around 10,000 annotated projects in different programming languages. For the classification, their approach uses the content of the \texttt{README} files and source code, represented using TFIDF, as input to a probabilistic model called Multinomial Na\"{i}ve Bayesian Network to recommend new possible topics for the project.

Similarly to~\cite{sipio2020naive}, Repologue~\cite{izadi2020topic} proposes a dataset based on popular GitHub Topics; however, the dataset is unavailable. For the classification, they also adopted a multimodal approach. They feed as input to a fully connected neural network, the dense vector representation (\ie embeddings) created by BERT~\cite{devlin-etal-2019-bert}, a neural language model. BERT creates the embedding of the project names, descriptions, \texttt{README}s, wiki pages, and file names concatenated.

GHTRec~\cite{9590294} has been proposed to recommend personalized trending repositories, \ie a list of most starred repositories, by relying on the BERT language model (LM) and GitHub Topics. Given a repository, the system predicts the list of topics using the preprocessed \texttt{README} content. Afterward, GHTRec infers the user's topic preferences from the historical data, \ie commits. The tool eventually suggests the most similar trending repositories by computing the similarity on the topic vectors, \ie cosine similarity and shared similarity between the developer and a trending repository. They use the dataset of \cite{sipio2020naive}.

\subsection{Automatic Taxonomy Construction}

Automatic taxonomy construction or induction is a challenging task in the field of natural language processing as it requires models understanding of the \textit{hypernym} relation. Hypernymy, or `\texttt{IS-A}' relation, is a lexical-semantic relation in natural languages, which associates general terms to their instances or subtypes.

With the large Web data available, many taxonomies are constructed from human resources such as Wikidata. However, even these huge taxonomies may lack domain-specific knowledge. Therefore, many automatic approaches to construct domain-specific ones have been proposed. From hypernymy discovery and lexical entailment~\cite{yu-etal-2020-hearst} approaches, to instance-based taxonomy~\cite{shang-etal-2020-taxonomy, chen-etal-2021-constructing},  and clustering-based taxonomy methods ~\cite{zhang2018taxogen, shang2020nettaxo}.

An example of approaches focusing on the hypernymy discovery task is ~\cite{yu-etal-2020-hearst}; they propose a distributional approach that fixes some of the issues present with such methods, making them achieve comparable performance with respect to the simple, pattern-based methods.

While shifting a bit from the hypernymy discovery methods, \cite{shang-etal-2020-taxonomy}, and \cite{chen-etal-2021-constructing} make use of the patterns matching for creating their datasets. In~\cite{chen-etal-2021-constructing}, the authors use a pre-trained language model and distantly annotated data collected by scraping the web. They finetune BERT to learn a hypernymy relation between words. In \cite{shang-etal-2020-taxonomy}, they use the dataset built by using pattern matching to construct a noisy graph of hypernymy and train a Graph Neural Network~\cite{scarselli2009gnn} using a set of taxonomies for some known domains. The learned model is then used to generate a taxonomy for a new unknown domain given a set of terms for the new domain.

Examples of clustering-based approaches include TaxoGen~\cite{zhang2018taxogen}, NetTaxo~\cite{shang2020nettaxo}, and Corel~\cite{huang2020corel}. Their work is similar in nature; all focus on creating a taxonomy from DBLP's bibliography, making it relevant to our research. However, attempts at reproducing their results have failed~\footnote{\href{https://github.com/xinyangz/NetTaxo/issues}{https://github.com/xinyangz/NetTaxo/issues}}\footnote{\href{https://github.com/teapot123/CoRel/issues}{https://github.com/teapot123/CoRel/issues}}, and their results are not publicly available, except for the small samples included in the paper. Their approaches are similar and are based on learning semantic vectors (embedding) for the words of interest. They perform an iterative sequence of learning embeddings: perform clustering and subsequently, for each cluster, repeat the steps to create each time a better representation that is more discriminative. However, this requires a large quantity of data, which is hard to collect~\cite{wang-etal-2017-short}; moreover, for each newly added term, a new run of the algorithms is required, which are heavily demanding in terms of computation and time. Furthermore, our attempts at reproducing their results failed. %, as the systems are stochastic.

A more comprehensive study of the taxonomy construction research area is presented in~\cite{wang-etal-2017-short}.

 \begin{figure*}[htb!]
    \centering
    \includegraphics[width=\textwidth]{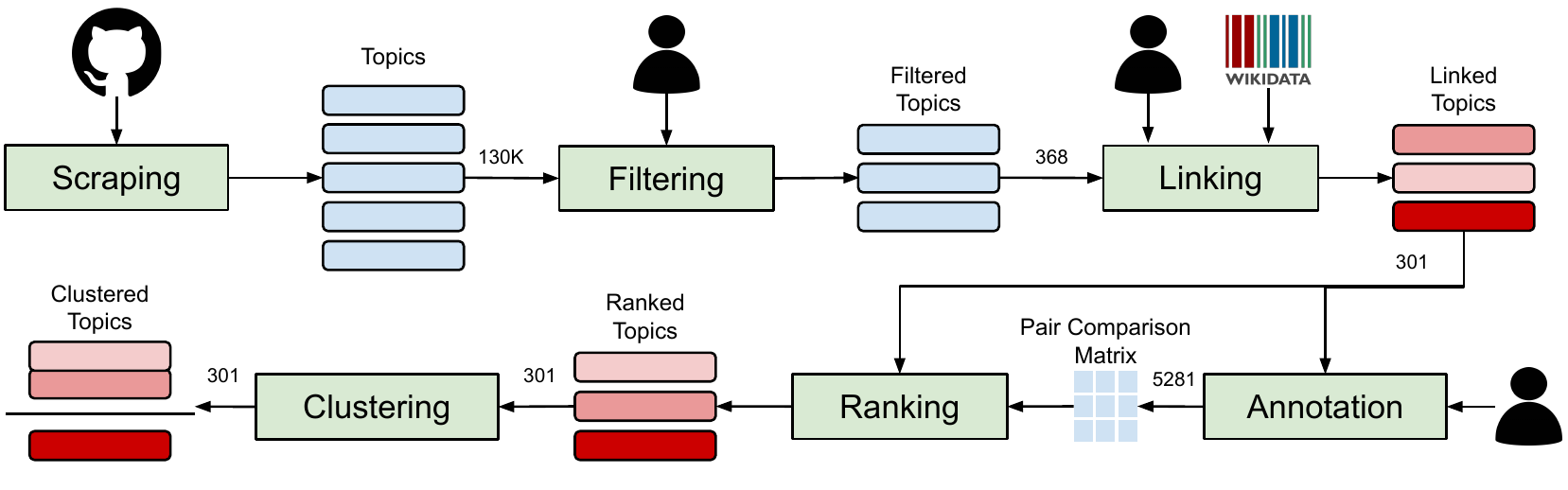}
    \caption{\NAME's pipeline for creating the ranking of GitHub Topics.}
    \label{fig:pipeline}
\end{figure*}

\section{Background}
\label{sec:_bkground}

Modeling subjective characteristics of items, \eg quality of an image or user preferences, requires subjective assessment and preference aggregation techniques to combine the human annotations. Usually, these consist of either a rating of a set of items based on some criteria or creating a ranking of a subset of the overall items. While the ranking is better suited for crowd-sourcing scenarios as it is less complex for the annotators~\cite{ye2014subjective}, compared to rating, it requires the inference of latent scores representing the position of the items in the rank, which involves comparison pairs samples.

The ranking task is defined as a comparison of $n$ items that are evaluated using subjective features without ground truth scores. The most straightforward experimental protocol is to compare pairs, referred to as pairwise comparison; however, this will take too many evaluations, more precisely $\binom{n}{2} = n(n-1)/2$. Nonetheless, \textit{active sampling} can be used to select the most informative pairs to compare, reducing the number of total comparisons while maintaining good results.

\subsection{Active  Sampling for Pairwise Comparisons}

\textit{Active  SAmpling for Pairwise comparisons} (ASAP)~\cite{mikhailiuk2020active} is a state-of-the-art active sampling algorithm based on information gain that finds the best pairs to compare in ranking experiments. 

Previous active sampling solutions reduce the computational complexity by taking a suboptimal approach of only updating the posterior distributions for the pairs selected for the subsequent comparison, which might not converge to the best optimum. Instead, ASAP reduces the overhead by using approximate message passing and only computes the information gain of the most informative pairs, updating the posterior distribution of all the pairs, making it efficient and correct.
%Compared to other active sampling algorithms \cite{}, which are suboptimal given that they update the posterior distribution only for the pairs selected for the next comparison, ASAP 

ASAP consists of two steps: \begin{enumerate*}[label=(\roman*)]
\item compute the posterior distribution of score variables $r$ using the pairwise comparisons collected; 
\item using the posterior of $r$ to estimate the next best comparisons to be performed.
\end{enumerate*}

The use of ASAP in this paper is to support the work of the annotators since this algorithm minimizes the number of comparisons needed to obtain a full classification.

% \subsection{Annotation}
% We use ASAP as an assist to our annotators, which will be presented with a two terms, and they have to pick which is the most general term between them, or if they are on the same level.

% %\cezar{Probably this should go in dataset, in the annotation part, we should focus a bit on the order of the sections}
% Using ASAP allows to reduce the amount of annotations required to achieve good performances. In the ASAP paper they show how reducing the the amount of comparisons on an example with 200 variable, affects the performances. In particular, a value of $\frac{1}{3}$ shows a good balance in terms of performances, and amount of comparisons, resulting in $\frac{1}{3}\binom{n}{2}$ total annotation. In our case, given the \textbf{B} terms we collected, and an estimated amount of 5 seconds required, we expect the overall time required to be $\frac{1}{3}\binom{\textbf{B}}{2} * 5 / 3600 = X$ hours. Given a pool of $K$ annotators, this will translate to $X/K$ hours for each annotators.

\subsection{Ranking Algorithm}
ASAP uses TrueSkill~\cite{herbrich2007trueskill} for the ranking of the annotated pairs. TrueSkill is a ranking system for calculating players' relative skills in zero-sum games. TrueSkill is similar to Elo~\cite{glickman1999parameter}, one of the first algorithm developed for ranking in two-player games. Elo models the probability of the possible game outcomes as a function of the two players' skills represented as a single scalar. However, unlike Elo, TrueSkill uses  Bayesian inference to evaluate a player's skill. Therefore, a player's skill is defined using a normal distribution $\mathcal{N}(\mu, \sigma)$, where $\mu$, the mean, is the perceived skill, and the variance $\sigma$ represents how uncertain the system is in the player's skill value. As such, $\mathcal{N}(x)$ can be interpreted as the probability that the player's ``true" skill is $x$.

TrueSkill, given its nature of being an online game ranking system, supports the addition of new players, or terms in our case, without needing to recompute pre-existing players' scores. Moreover, for pairwise comparisons, TrueSkill should be able to place the newly added element with around 12 comparisons\footnote{\href{https://trueskill.org/\#rating-the-model-for-skill}{https://trueskill.org/\#rating-the-model-for-skill}}. We validate this for our case in the next section.

%Lastly, we discretize the ranking space using Gaussian Mixture Model, the optimal number of clusters is found using the Silhouette coefficient. 

\section{Proposed Approach}
\label{sec:_approach}
\NAME  is our proposed approach for generating a hierarchical taxonomy of software application domains. It is a bottom-up ranking based on GitHub Topics and grounded in Wikidata. It aims to solve some of the issues present in current datasets for software classification, including: \textit{mixed taxonomies}, \textit{mixed granularity}, and \textit{ambiguity}.

In this section, we present the pipeline used to create the ranking of the GitHub Topics. The pipeline is visually represented in Figure~\ref{fig:pipeline}, and its activities are described in more detail below.

\subsection{Topic Collection - Scraping}
We collected the GitHub Topics by following the approach used in~\cite{izadi201repologue}. We scraped repositories containing (1) at least one GitHub Topic, (2) a \texttt{README} file, (3) a description, and (4) at least ten stars. In this way, we were able to retrieve 135K projects, with a total of 121K different topics, with a combined frequency of 1 million. We have made the list and metadata of the scraped projects available. Our data follows the format of GitHub's REST API\footnote{\href{https://docs.github.com/en/rest/reference/repos\#get}{https://docs.github.com/en/rest/reference/repos\#get}}.  

\begin{figure}
    \centering
    \includegraphics[width=\linewidth]{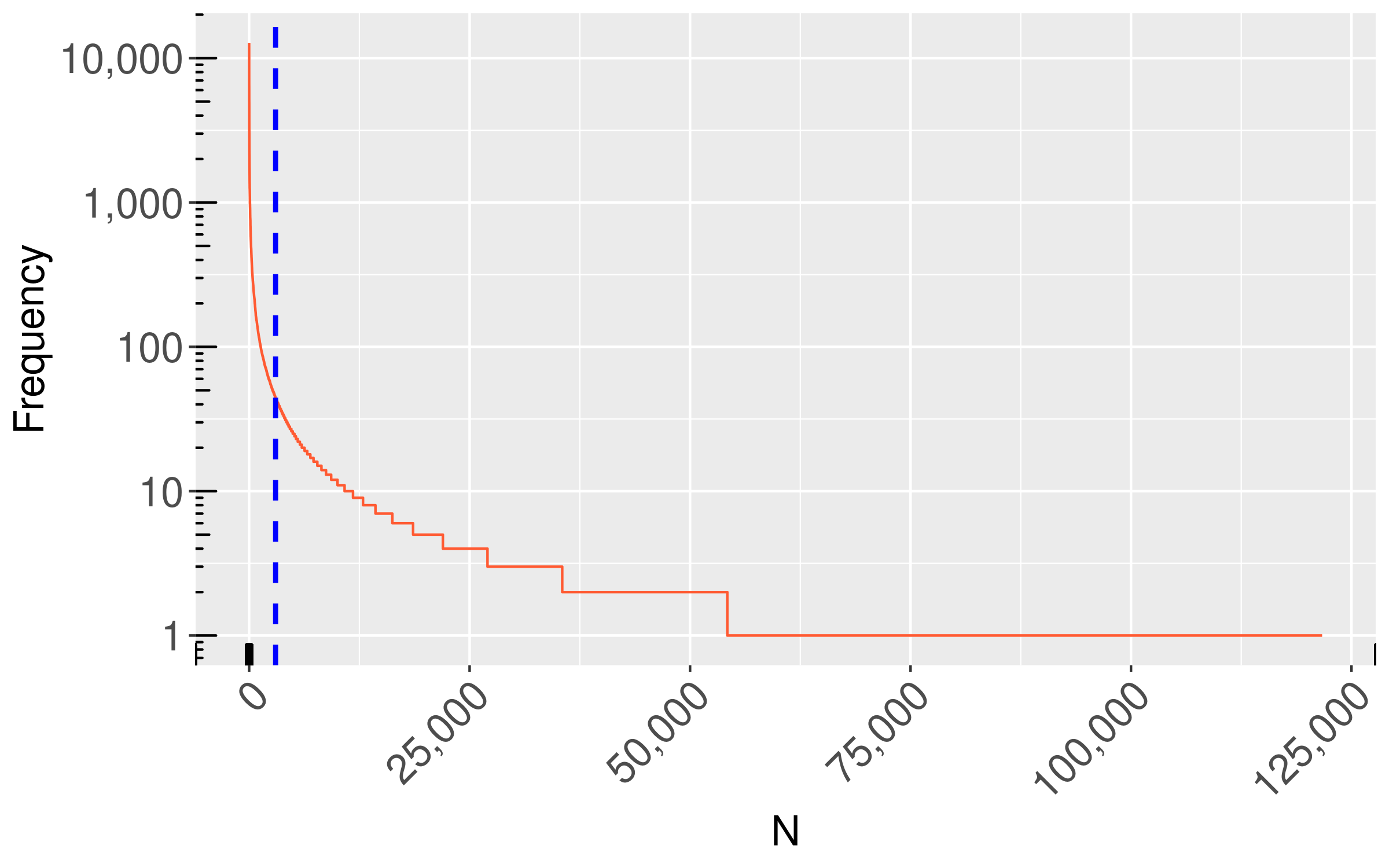}
    \caption{Distribution (log scale) of the frequency of the scraped topics from GitHub. The blue dotted line represents the cut line we picked for our initial filtering at $N=3,000$.}
    \label{fig:topic_dist}
\end{figure}

The distribution of the frequency and usage of the topics is highly skewed, as depicted in Figure~\ref{fig:topic_dist}, with around 50\% of projects annotated with only one topic. Less than 2,500 topics represent 50\% of the distribution of use. This is caused by many observable factors, including the popularity of specific programming trends (\eg `\textit{Machine Learning}'), and the usage of programming languages as topics (15 programming languages are listed in the top 50 topics, covering 7\% of the overall topic frequency). 

\subsection{Topic Filtering}
In this activity, we attempted to reduce the number of topics by filtering and manual annotation. Given the large variety of terms, an automatic approach based on the semantics of the topics would be preferred. However, given the absence of a precise context, the ambiguity of the task and terms, this approach is not optimal and will produce poor results. Therefore, we opted for manually annotating a subset of the topics as a solution.

For the annotation, we selected the 3,000 most frequent topics. Their frequency covers 60\% of the topics scraped in the previous step; each of those topics has a number of examples close to 50 (with a minimum of 44), meaning that there is enough data to train a machine learning model. %The final datasets stats are presented in Section~\ref{}. 

With the help of three annotators, we assigned a binary label (0, 1) to each of the 3,000 topics. The annotators were instructed to positively label (\ie 1) the GitHub Topics that can be considered as general or specific application domains of software (\eg `\textit{deep learning}', `\textit{common line interface}', etc.). At the same time, for programming languages, companies, technology, and any other case, the annotators were instructed to assign a null label (\ie 0).

As the final step of this activity, the selection of the resulting topics was carried out using a majority voting (\eg at least two annotators agreeing on a topic being an application domain); this was done to ensure higher quality in the creation of the initial taxonomy and to remove noise, while still allowing for a good recall. This activity filtered an overall 368 topics that can be considered application domains, and that were carried forward to the next reconciliation activity below.

\subsection{Topic Linking}
The topics resulting from the filtering and the manual annotations are intrinsically ambiguous (\eg `\textit{rna-seq}', or `\textit{ci}'); in order to help with the disambiguation of these terms, we linked each of the topics to Wikidata~\cite{vrandecic2012wikidata}, Wikimedia Foundation's knowledge base. The linking is performed in a semi-automatic fashion using Wikidata reconciliation API and humans to check and fix any errors. % in the process.

Wikidata offers a reconciliation API, a service that, given a text representing a name or label for an entity, and optionally additional information to narrow down and refine the search to entities, returns a ranked list of potential entities matching the input text (\eg for `\textit{rna-seq}' returning `\textit{RNA sequencing}', with Wikidata ID \texttt{Q2542347}\footnote{\href{https://www.wikidata.org/wiki/Q2542347}{https://www.wikidata.org/wiki/Q2542347}}, and for `\textit{ci}', will return the `\textit{Continuous Integration}' entity with ID \texttt{Q965769}). The reconciliation uses fuzzy matching to find the most likely entity in the knowledge base that matches the input string. Hence, the candidate text does not have to match each entity's name perfectly, meaning that we can go from ambiguous text names to precisely identified entities in a knowledge base.

The topics resulting from the previous activity were fed as an input to the Wikidata API: in order to increase the retrieval precision, we exploit the \textit{github-topic} Wikidata property, with ID \texttt{P9100}\footnote{\href{https://www.wikidata.org/wiki/Property:P9100}{https://www.wikidata.org/wiki/Property:P9100}}, that helps in the linking of terms that are already linked in Wikidata (\eg the entity `\textit{Convolutional Neural Network}', has an entry with property \texttt{P9000}, where the value is `\textit{convolutional-neural-network}'). 

%This reconciliation activity gives us a list of 10 reconciliation candidates, for each term. Each candidate has a list of types describing the candidate. We manually annotate the highly irrelevant types to the task (\eg \textit{human}, \textit{scholarly article}, or any location) and exclude candidates belonging to these types for a more automated process of linking (\hl{this is really not clear, and misses some important info and details to understand}). Lastly, we check the correctness of the linking and fix improperly linked topics. This resulted in the correction of 25 topics out of the 368. 

This reconciliation activity gives us a list of 10 candidates for each term. Each candidate has a list of types describing the candidate (\eg for \textit{`Science'}, there are various candidates with the same name, but different types, some include: `\textit{academic discipline}', which would link to the correct entity \texttt{Q336}, and the other will link to `\textit{television channel}' with ID \texttt{Q845056}). We manually annotate the highly irrelevant types to the task (\eg \textit{human}, \textit{television channel}, or any location) and exclude candidates belonging to these types for a more automated process of linking. After filtering the candidates that are of an irrelevant type, we link the term by picking the first candidate in the filtered list. Lastly, we check the correctness of the linking and fix improperly linked topics. This resulted in the correction of 25 topics out of the 368.

%Lastly, using the resulting ID from the reconciliation, we combine the topics that point to the same entity, resulting in a total of 301 topics that we will use to create our taxonomy.

Now that our topics are disambiguated, we can use the unique ID from Wikidata to reduce duplicates, as some topics are just different surface forms, or aliases, for the same entity. The number of unique topics remained are 301.

%(\hl{this step needs some rewriting work, especially to discuss why it is needed in the first place})

\subsection{Annotation}

\begin{figure}[htb!]
    \centering
    \includegraphics[width=0.75\linewidth]{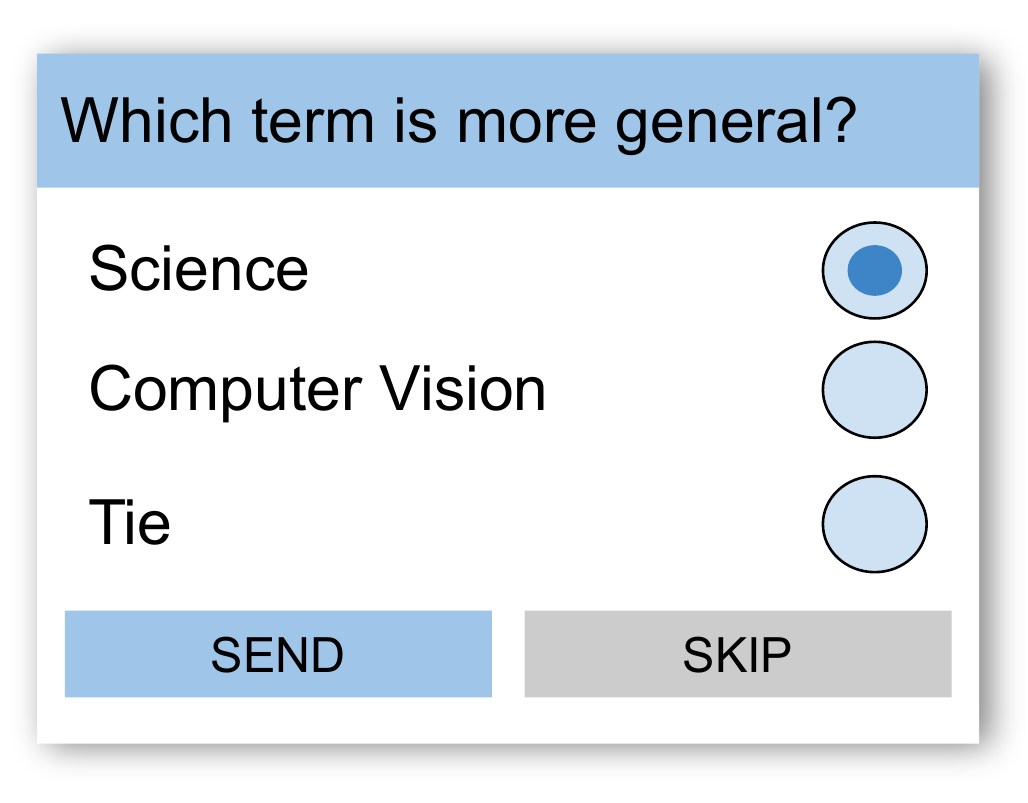}
  \caption{The user interface presented to the annotators. The skip bottom allows the annotators to obtain a new pair when they are not confident with the current.}
    \label{fig:choice_ui}
\end{figure}

In this activity the resulting topics (filtered by the annotators and linked to the Wikidata API) were presented in pairs on a web application for the manual annotation. The $8$ annotators working on this activity were presented with two topics in the list and were instructed to pick \textit{the most general term}, considering their respective domain. A mock up of the interface is illustrated in Figure~\ref{fig:choice_ui}. The terms were also linked to the URLs of the Wikidata pages, in case the annotators were not confident with a specific topic. In case of doubt, the annotators could `skip' the pair. 

The \textit{`Tie'} option was also available, in case the annotators believe that the two terms represent domains of the same level. This option was also used to collect data to validate our results: since ASAP does not support  \textit{`Ties'}, we instructed the annotators to use it rarely. Instead, annotators were instructed to rather pick one of the options randomly, as having one term before the other in the continuous ranking does not affect the final discrete rank.

The ASAP algorithm was used to assist the work of the annotators: its main advantage is to reduce the number of pairwise annotations, and still achieve decent performance. The ASAP paper~\cite{mikhailiuk2020active} shows in fact how reducing the number of comparisons on an example with 200 variables affects the performances. In their case, a reduction in the number of comparisons by a factor of ${3}$ shows a good balance in terms of performance, resulting in a total of $\frac{1}{3}\binom{n}{2}$ annotations needed. 

The ASAP algorithm identifies the topic pairs, and it uses the previous annotation to find the best, most informative, new pairs to add to the list of annotations. The annotators are then presented with a random pair from this pool. The ASAP algorithm is a very memory-intensive task; we used an AWS instance with 64 GB of RAM and a 16 core CPU for our experiments.

In terms of performance of our annotation, and given the 301 terms we collected and an estimated amount of 25 seconds per pair on average, we expected the overall time required to be $\frac{1}{3}\binom{301}{2} * 25 / 3,600 = 104$ hours. Given our pool of $8$ annotators, this would translate to $13$ hours for each annotator. However, we would expect fewer samples required for our case, as the task is less challenging compared to modeling more abstract values like a player skill in an online game. % \hl{unclear, why less challenging?}. - CHANGED

\subsection{Ranking}

The final ranking of the topics is computed by ASAP using TrueSkill. It uses the comparison matrix created with the annotations at the previous step, excluding the items marked with \textit{`Tie'}. The comparison matrix is a square matrix, where every entry is the number of times the term in the corresponding row was selected over the term in the corresponding column. This results in a mean and a standard deviation value for each topic, and by sorting by the mean of each topic, we obtain the ranking. 

For example, given the terms: `\textit{Science}', `\textit{Computer Science}', and `\textit{Software Engineering}', and a comparison matrix, the final ranking returned by TrueSkill will be:

\begin{itemize}
    \item \textit{Science}: $\mathcal{N}(2, 0.3)$;
    \item \textit{Computer Science}: $\mathcal{N}(1.5, 0.3)$;
     \item \textit{Software Engineering}: $\mathcal{N}(0.9, 0.3)$.
\end{itemize}

%\hl{add a snippet of the resulting ranking, for clarity}.

\subsection{Clustering}

Lastly, the final step of our pipeline is to create a discrete rank of the topics. Using the resulting ranking of ASAP, we feed the mean of the topics ranking as input to a clustering algorithm, KMeans. To find the optimal number of clusters, we used the Elbow method. The clustering is performed on the uni-dimensional data of the topics' mean score computed by TrueSkill.

Having a discrete set of ranks, instead of a continuous one, brings many possibilities: perform analysis to study developers behaviour, and use the discrete values to train models to predict topics at specific levels, aiding with the annotation of repository.

% We furthermore confirmed our results by and also using BIC method, however it only works for the soft clustering version of K means, Gaussian Mixture Models (GMM).

\subsection{Ranking New Topics}
\label{sec:new_topic}
We evaluated the number of annotations required to rank newly added topics. The experiment uses the annotations collected from the experiments with annotators. We simulate a new topic addition by removing their annotations and incrementally adding them, one by one, and checking for convergence. We measure the average number of annotations required to reach convergence for all 301 topics by individually removing each one.

We compare three different strategies of simulating the newly inserted topic: \textbf{random}, \textbf{order}, and \textbf{informed}. The  \textbf{random} strategy samples annotations of the topic randomly; \textbf{order} selects the pairs in order as annotated and suggested by ASAP; the \textbf{informed} uses only the last 20 pairs suggested by ASAP, making it a more efficient, but not optimal way, to simulate the new annotations.

%We compare three different strategies of annotating the newly inserted topic: \textbf{random}, \textbf{order}, and \textbf{informed}. Where the \textbf{random} samples annotations of the topic randomly, \textbf{order} selects the pairs in order as annotated, the \textbf{informed} uses the last 20 annotations as extracted by ASAP, making it a more efficient, but not optimal way, to simulate the new annotations.

The convergence is defined as being in proximity of the final position that the topic holds in the ranking used with all the annotations at our disposal. We use a max of 3 positions difference for proximity, and convergence is reached when the proximity is hold for 2 consecutive annotations. 

\section{Results}
\label{sec:_results}

In this section, we present the results of our approach and discuss them. We present the statistics of the filtering and annotation. We also present the ranking and samples from it. Lastly, we show the results of adding a new term to the ranking.

\subsection{Filtering}
Starting from the 3,000 topics, covering 60\% of the total topics distribution, the final list selected by the annotators contains 368 topics. 

Table~\ref{tab:num_positive_ann} shows the number of positively labelled topics from each annotator, with their positive rate. The three annotators had different ideas of what an application domain is. Annotator \textbf{C} was more strict, picking a minimal amount of terms, \textbf{A} was  more relaxed picking many, with \textbf{B} in between. Furthermore, we measured the inter-rater reliability using Krippendorff's alpha~\cite{krippendorff1970estimating}, a general, reliable measure for inter-rater reliability~\cite{krippendorff2004reliability} suited for any number of annotators. For the 3,000 topics and our three annotators, we obtained an agreement score of $0.68$. If we measure it per pair of annotators, we have an agreement of $0.79$ for pair \textbf{A-B}, $0.68$ for the \textbf{B-C} one, and $0.52$ for pair \textbf{A-C}.

The low amount of GitHub Topics that qualify as application domains suggests that using popularity as seed for a taxonomy, which previous work do, results in low quality labels. 

\begin{table}[htb!]
\caption{Number of positive labelled topics for each annotator and their positive rate.}
\label{tab:num_positive_ann}
\begin{tabular}{ccc}
\toprule
\multicolumn{1}{c}{\textbf{Annotator ID}} & \multicolumn{1}{c}{\textbf{Positive}} & \multicolumn{1}{l}{\textbf{Positive Rate}} \\
\midrule 
A                             & 496                          & 0.16            \\
B                             & 370                          & 0.12            \\
C                             & 238                          & 0.08            \\
\bottomrule
\end{tabular}
\end{table}

\begin{mybox}[]
Using 60\% of the most frequent topics as seed allows us to reduce bias in the representation of less common topic. The filtering avoids the presence of terms that are not application domains (\eg programming languages). Both issues are common in previous work. 
\end{mybox}

\subsection{Annotation}

From the pairwise comparison data annotation, we collected 5281 annotated pairs from 8 annotators, including Professors, PostDocs, and PhDs in Computer Science with a mix of Software Engineering and Machine Learning backgrounds. The statistics about the annotators' contribution to the process is presented in Table ~\ref{tab:annotation_stats}.

\begin{table}[htb!]
    \caption{List of the annotators' IDs and their contribution to the total annotations with the number of ties assigned.}
\centering
\begin{tabular}{ccc}
\toprule
\textbf{Annotator ID} & \textbf{Annotations} & \textbf{Ties}   \\
\midrule
1          & 1,207   & 117 \\
2          & 454    & 43 \\
3          & 1,520   & 11 \\
4          & 94     & 0 \\
7          & 561    & 82 \\
8          & 246    & 27 \\
9          & 764    & 89 \\
10         & 190    & 13\\
\midrule
\textbf{Total} & 5,281 & 382  \\
\bottomrule
\label{tab:annotation_stats}
\end{tabular}
\end{table}

\begin{figure}[htb!]
    \centering
    \includegraphics[width=\linewidth]{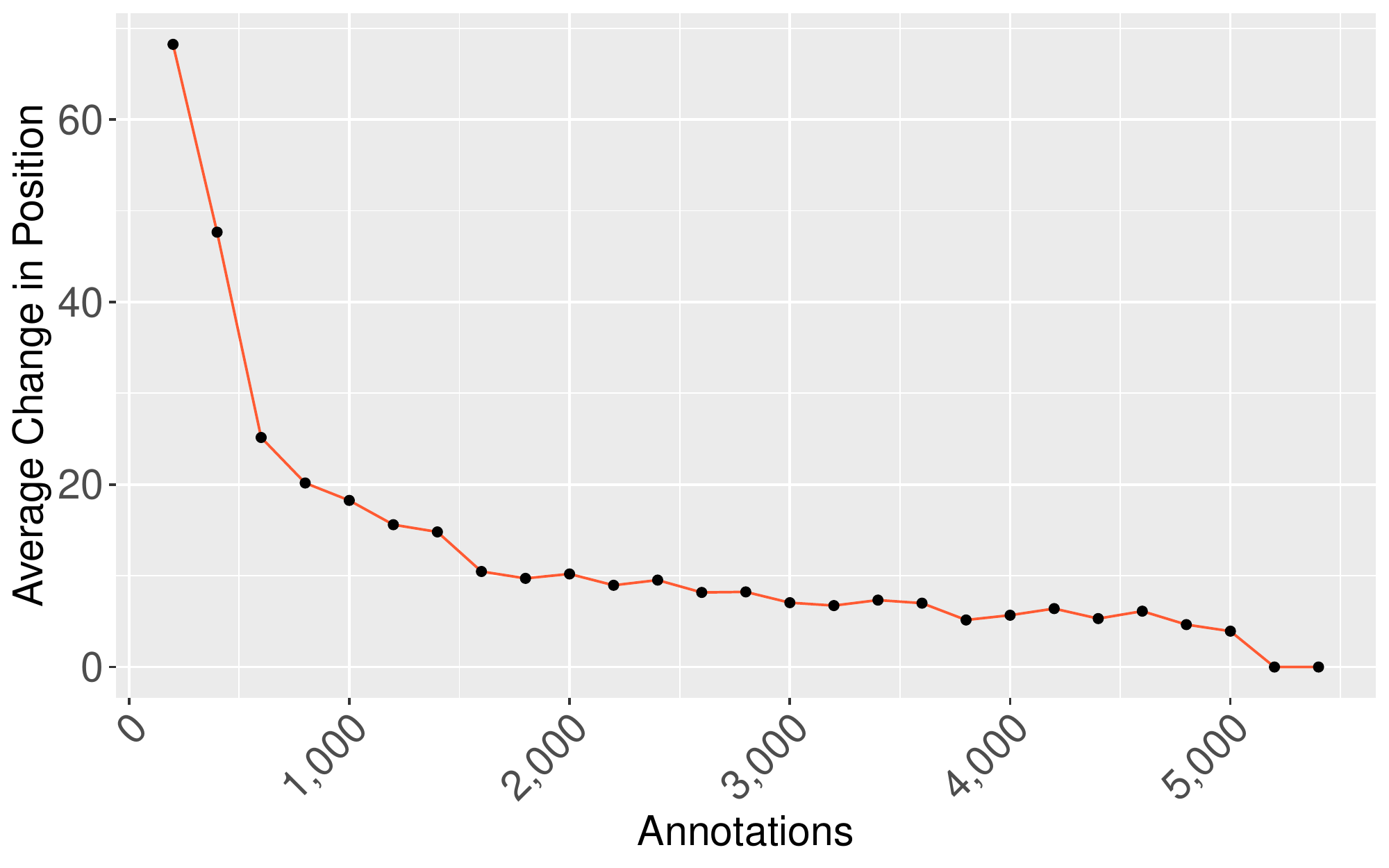}
    \caption{Average change in position of the terms every 200 new annotations.}
    \label{fig:convergence}
\end{figure}

As we predicted, we were able to converge to a stable rank in less than the $\frac{1}{3}\binom{301}{2} = 15,050$ annotations that we would have expected, as based on the case study in the ASAP paper. From Figure~\ref{fig:convergence}, which shows the average change in position at incremental amounts of annotation, we can see that the average change in positions in the ranking converged at around $\frac{1}{9}\binom{301}{2} = 5,000$ annotations. This also means that there is no further change in the positions of the elements in the ranking. 

The curve has a steep decrease in the first 1,000 comparisons, and later plateaus at an average of 10 positions changed by the terms in the ranking. After 5,000, it immediately falls to an average close to 0. 

\begin{mybox}
ASAP reduces the amount of annotation required to reach convergence in the ranking by a factor of 9, making it an effective way to aggregate domain expertise in qualitative ranking tasks.
\end{mybox}

\subsection{Ranking}

\begin{figure}[htb!]
    \centering
    \includegraphics[width=\linewidth]{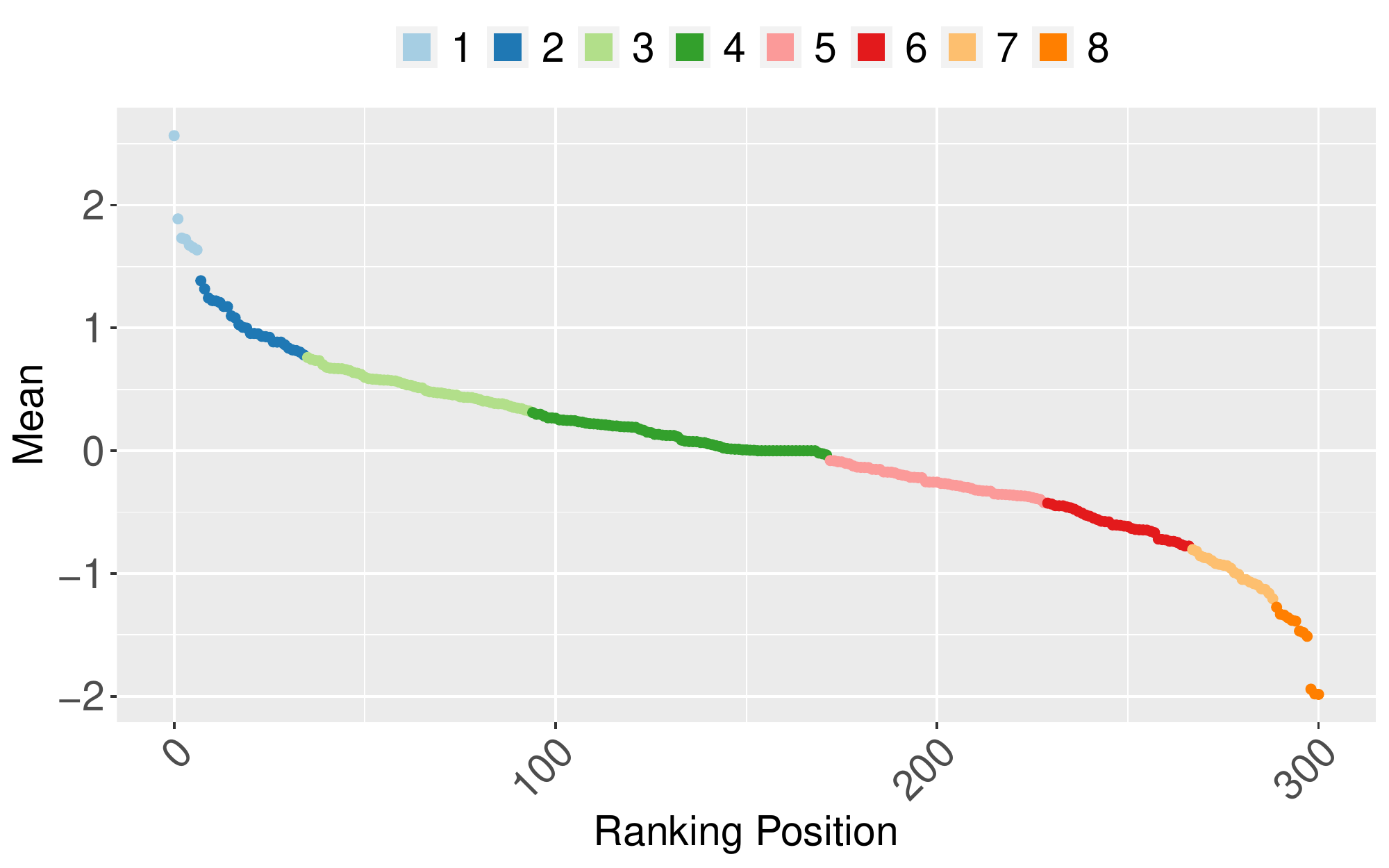}
    \caption{Ranking of the GitHub Topics. The $x$-axis represents the position, while the $y$-axis is the mean score extracted by TrueSkill. The color represents the cluster that each topic is assigned to.}
    \label{fig:scatter_ranking}
\end{figure}

The final ranking is presented in Figure~\ref{fig:scatter_ranking}, where we see the topics' mean score computed by TrueSkill, and their position in the final rank. We can notice that, while the extremity of the ranking is very well separated, the central area is not as much. This is caused by the higher difficulty of comparing topics belonging to the middle area of a taxonomy. 

A more qualitatively view of the resulting ranking is presented as a sample of the topics at different levels in Table~\ref{tab:topic_rank_list}. The Table shows a vertical view of topics that would belong to the same branch in a taxonomy. In particular, it presents the Artificial Intelligence / Computer Vision branch, from the top most general term to middle terms, all the way to the last terms in such a branch. From Table~\ref{tab:topic_rank_list}, we notice that at the top we have terms that we would come to expect with `\textit{Science}' being first one. As we go down the ranking, the terms get more specific, first `\textit{Computer Science}' followed by `\textit{Artificial Intelligence}'. After these, we find a limit case, where we have two terms `\textit{Computer Vision}' and `\textit{Machine Learning}' that for someone might need to be reversed; for someone, they are correct, and for others, they should be at the same level. As we get closer to the end, we find more concrete tasks like `\textit{Image Segmentation}'. At the bottom, we see methods like `\textit{Convolutional Neural Network}'. 

The final list can be checked in the data replication package (see Section~\ref{sec:intro}).

\begin{table}[htb!]
    \caption{Rank of a subset of terms in the vertical of Computer Science, Machine Learning, and Computer Vision.}
\centering
\begin{tabular}{cc}
\toprule
\textbf{Rank} & \textbf{Topic}  \\
\midrule
1                       & Science  \\
2                       & Mathematics   \\
3                       & Physics  \\
4                       & Engineering    \\
\textbf{\dots}          & \textbf{\dots} \\
9                       & Computer Science   \\
\textbf{\dots}          & \textbf{\dots} \\
34                      & Artificial Intelligence   \\
\textbf{\dots}          & \textbf{\dots} \\
60                      & Computer Vision \\
61                      & Machine Learning \\
\textbf{\dots}          & \textbf{\dots} \\
147                     & Image Segmentation\\
\textbf{\dots}          & \textbf{\dots} \\
295                     & Convolutional Neural Networks \\
\textbf{\dots}          & \textbf{\dots} \\
\bottomrule
\label{tab:topic_rank_list}
\end{tabular}
\end{table}

\begin{mybox}
TrueSkill's ranking captures the hierarchical relations among terms. However, there is still room for improvement at the middle levels as the separation is not as strong.
\end{mybox}

\subsection{Clustering}

Using the Elbow method for KMeans, we found the optimal number of clusters for our ranking at $n=8$. In Figure~\ref{fig:scatter_ranking}, we can see the topics ranking distribution and their cluster.

We evaluate our clustering using the \textit{`Tie'} labelled pairs from the annotation phase. We measure how many of the \textit{`Tie'} annotated pairs end up in the same buck in the cluster. Out of the 382 ties, 364 were unique, almost evenly distributed among the eight categories. The results show 30\% of the tied pairs belong to the same cluster. When loosening the constraint of equality and setting a distance of 1 cluster, we reach 100\% accuracy of our clustering.
This result suggests that overall our method is effective; nevertheless, many cases are not precisely placed at the correct level. However, if we also take into account the results in Table~\ref{tab:topic_rank_list}, we can see that terms in the same vertical are correctly ordered. Still, across verticals, there might be less order, which is in line with the objective of the ranking: \textit{create a sorting of terms that belong to the same domain}.

\subsection{Topic Ranking Distribution}

\begin{figure}[htb!]
    \centering
    \includegraphics[width=\linewidth]{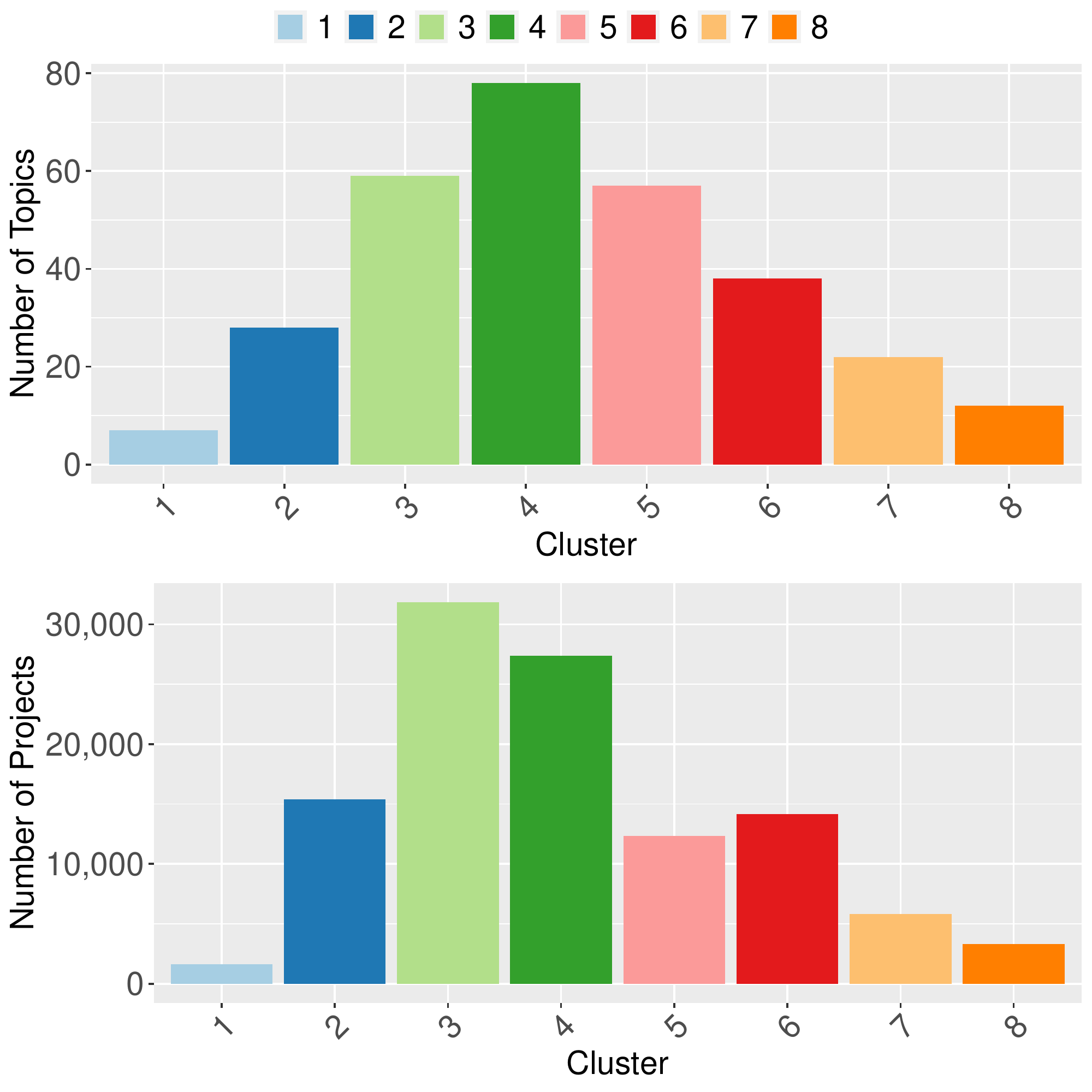}
    \caption{(Top) Distribution of the topics in the clusters. (Bottom) The number of projects (frequency of the topics) in each cluster. A low cluster number means more general, and a higher value means more specific.}
    %\caption{Distribution of the number of projects (based on the topics it is labeled with) in the extracted ranking. A low cluster number means more general, a higher value means more specific.}
    \label{fig:cluster_dist}
\end{figure}

After obtaining our ranking, we are able to answer: 

\smallskip
\begin{quote}
    \textit{\textbf{RQ}: Are the topics used to annotate GitHub projects evenly distributed in the levels of a taxonomy?}
\end{quote}
\smallskip
To answer this, we can use Figure~\ref{fig:cluster_dist}, which shows the distribution of topics at different levels in the ranking. The top bar chart shows that the topics are normally distributed across the levels, with the mean at level four, right in the middle. However, if we take into account the frequency of the topics (bottom plot), the mean moves towards a higher, and more general level, level three. 

This suggests a tendency on the developers in using a general term that only describes the area of application but without specific information. This has a negative impact on the retrieval of projects, and affects negatively the time required to find the appropriate repository, as not many are labelled with a more specific term.

\begin{mybox}
There is a lack of specificity in the terms used by developers to annotate their projects. Future work in software classification needs to address this issue by suggesting topics at multiple levels.
\end{mybox}

\subsection{Ranking New Topics}
% The results of measuring the number of annotations required for adding a new topic in the ranking are presented in Table~\ref{tab:new_topic}. We can notice that, if we use a non random sampler, the convergence of the topic requires a small amount of topic, around 15, which is a bit more than the one suggested by TrueSkill. This shows how easy is to extend the classification to keep it up to date, or adapt it to new sub domains. For batch additions, the required amount of annotation scales linearly with the number of new topics as the problem can be viewed as multiple single term increments.

Using the different approaches to simulate annotations for a newly added topic defined in Section~\ref{sec:new_topic} (`random', `order' and `informed'), we measured the average amount of annotations required to reach convergence. 

The results %of measuring the number of annotations required for adding a new topic in the ranking 
are presented in Table~\ref{tab:new_topic}. We can notice that, independently of the scenario, the amount of annotations required is minimal, with a range from 22, in a non-optimized scenario, to around 15 when using a better approach (e.g., `informed'). These are more than the ones suggested by TrueSkill. However, the difference is negligible if we consider that the selection of the pairs was not online, as it would be for new topics.

This shows how easy it is to extend the classification to keep it up to date or adapt it to new subdomains. Furthermore, the required amount of annotation scales linearly with the number of new topics for batch additions, as the problem can be viewed as multiple single-term increments.

\begin{table}[htb!]
    \centering
    \caption{Average number of annotations required to reach convergence when adding a new term in the ranking. }
    \begin{tabular}{cc}
        \toprule
        \textbf{Method} & \textbf{Average}  \\
        \midrule
        random  & 22  \\
        order & 22  \\
        informed & 15 \\
        \bottomrule
    \end{tabular}
    \label{tab:new_topic}
\end{table}

\begin{mybox}
Our pipeline is flexible and allows for the insertion of new terms with a minimum effort, making our taxonomy a good starting point to build upon.
\end{mybox}

\section{Discussion}
\label{sec_discussion}
In this section, we discuss what are the unique features, in our opinion, of the framework that we have presented. We also discuss the repercussions of the choices made and what implications should be expected.
% \begin{itemize}
%     \item how to add more terms
%     \item implications of using and reducing `only 3,000' terms
%     \item how could this be relevant for better GitHub annotations
%     \item top and bottom of the ranking are clearly separated, but how to make the middle part more clearly separated (using future work?)
% \end{itemize}
\paragraph{Unique features} Our exploration of a bottom-up, data driven taxonomy has shown that the labels used by developers in their everyday work can form a solid starting point for a taxonomy. Although this has been attempted in the past, we believe that this works adds at least two unique features to this quest: first, the approach that we developed is based on a \textit{seed} that was annotated by $8$ experts, and whose provenance is directly rooted in the development of thousands of active GitHub projects. The annotation part, albeit time-consuming and process-intensive, is a necessary factor for the quality of the seed: this activity has been mostly absent in all the past works that we analyzed for reference.

\paragraph{Extensibility} The second unique feature that this work offers to the research community is a flexible and dynamic approach to \textit{expand} the taxonomy to further terms and labels. All existing taxonomies can be considered as flexible, in the sense that they allow for further terms to be included. The added value in the work that we propose is that \NAME allows to dynamically allocate a new label by means of further annotations by anyone proposing such a label. For example, if a researcher wanted to add a new label, not previously present in our taxonomy, they would be expected to run some 15 pairwise comparisons. That would be necessary and sufficient to locate the new term in the correct place of our taxonomy. The ability to add new terms is crucial, as the field evolves and new terms might become popular quickly, or someone wants to adapt it with some terms of interest that are not currently present.

\paragraph{Initial seed} For our study we decided to use as a starting point  the top 3,000 topics by frequency. This sample covers 60\% of all the labels present in the overall set of 135K projects, and these labels do not include less popular or underrepresented topics. One might argue that this was the reason of our results regarding the distribution of the topics in the levels: the less popular are also the more specific terms. However, we perform the ranking and clustering on the topics available, hence the results would be even more skewed towards higher levels, if we considered the less frequent topics as well. Furthermore, the further down the list we go, the more noise and duplicates we find, making it more time consuming for the annotators.

%Moreover, one key feature of our framework is the ability to dynamically add new terms to the ranking. This is aided by TrueSkill, as it is designed for online games. The addition can be either performed with the support of ASAP to find the best candidate, or by iteratively comparing the term with a term in the neighbourhood and a higher mean score, annotate  the pair, and then update the score and repeat. 

\paragraph{Practical applications of the taxonomy} Taxonomies and classifications have an inherent utility in organizing the knowledge base around a specific area of expertise. In our case, we believe that having such a classification can be further used by GitHub to guide the developers in labelling their project with at least one term for each level. This would be similar to the description of an academic paper using the ACM Computing Classification System, that allows research to choose from high-level and lower-level topics to describe their work. Lastly, the taxonomy can be used to automatically suggest topics at all levels in the ranking to repositories on GitHub, improving retrievability.

\paragraph{Improvements} While our framework has shown its ability to create a good ranking of the selected terms, in Figure~\ref{fig:scatter_ranking}, as mentioned in the Results section, the separation at the middle layer is not as strong as we would like. This could be addressed by collecting more pairwise comparison data for only the middle area, however, contrary to the expectation, this might not be the case as the middle layer is also the hardest to separate for humans. One solution could be to perform linking among terms, making a tree like taxonomy. However, this is also not as trivial, and requires more research.

\section{Threats to Validity}
\label{sec:_threats}
We use the classification of Runeson et al. \cite{Runeson2012threats} for analyzing the threats to validity in our work.  We will present the \emph{construct validity}, \emph{external validity}, and \emph{reliability}. Internal validity was not considered as we did not examine causal relations \cite{Runeson2012threats}.

\subsection{Construct Validity}
The first construction threat to our study is the initial filtering of the topics, which is highly subjective. However, we reduced this threat by having three annotators and only choosing the topics agreed on by at least two of them. For the ranking, while we have subjective input, the algorithm combines the input from all annotators, reducing the weight of annotation errors.

Furthermore, our ranking is open to evolution, by having a flexible pipeline, additions and changes can be done with a minimal effort, which can be open to the community.

Regarding the analysis on the distribution of the ranking and how developers label their projects, we mitigated this thread by using the validation data collected from the annotation phase.

\subsection{External Validity}
Our approach is independent of the domain, as it uses general methodologies from statistics and machine learning. Furthermore, we focus on words, making this approach applicable to all domains, not just software application domain ranking.

\subsection{Reliability}
For the initial selection of projects, we collected a high amount of different projects, resulting in a large pool of terms, making the collected pool a good sample for representing the population. Regarding the filtering, as discussed in the Construction Validity section, working with natural text is inherently subjective; we focused on having robust filtering of the topics used for our study. 

\section{Conclusions and Future Work}
\label{sec:_conclusion}

This paper presented \NAME, a framework for creating a discrete ranking of software application domains. Our work aims at solving some of the common issues present in current datasets for software classification, including: \textit{mixed taxonomies}, \textit{mixed granularity}, and \textit{ambiguity}. 

Using \NAME, we analyzed the top 60\% of a large sample of GitHub Topics, and selected a list of 301 that we considered application domains. We then disambiguated each topic by linking them to the Wikidata knowledge base. Furthermore, aided by the ASAP active sampling algorithm, 8 annotators compared more than 5,000 topic pairs: finally the TrueSkill algorithm used those annotated pairs to create a ranking of the selected application domains. \NAME's pipeline allows the resolution of the previous issues.

As the last contribution, we answered our research question \textbf{RQ}: by performing clustering of our ranking, we were able to find that developers tend to assign high-level labels to their projects, making it harder to find specific projects. \NAME proves as a viable option for developers to annotate their projects with more specific terms.

%As a solution, we suggest that developers keep into account the taxonomic level of the terms they use to annotate their projects. 

%\subsection{Future Work}
We plan to improve on our work in different ways: first, we would like to increase the number of topics in the list and ranking. Furthermore, we would create a hierarchical taxonomy and link the terms in our ranking. Moreover, we are interested in creating mappings for the terms in the lower end of the distribution, as there are many surface variant of topics already present in our taxonomy. This will allow for automatic, distant annotation of GitHub projects, which translates in the creation of a large-scale multi-label dataset for software classification that can evolve. Lastly, we plan to train classification models that are able to automatically recommend topics at specific levels, which will make it easier for developers to properly label their projects. 
\bibliographystyle{ACM-Reference-Format}
\bibliography{references}


\begin{thebibliography}{39}

%%% the Web address.

\ifx \showCODEN    \undefined \def \showCODEN     #1{\unskip}     \fi
\ifx \showDOI      \undefined \def \showDOI       #1{#1}\fi
\ifx \showISBNx    \undefined \def \showISBNx     #1{\unskip}     \fi
\ifx \showISBNxiii \undefined \def \showISBNxiii  #1{\unskip}     \fi
\ifx \showISSN     \undefined \def \showISSN      #1{\unskip}     \fi
\ifx \showLCCN     \undefined \def \showLCCN      #1{\unskip}     \fi
\ifx \shownote     \undefined \def \shownote      #1{#1}          \fi
\ifx \showarticletitle \undefined \def \showarticletitle #1{#1}   \fi
\ifx \showURL      \undefined \def \showURL       {\relax}        \fi
% invisible to TeX
\providecommand\bibfield[2]{#2}
\providecommand\bibinfo[2]{#2}
\providecommand\natexlab[1]{#1}
\providecommand\showeprint[2][]{arXiv:#2}

\bibitem[Altarawy et~al\mbox{.}(2018)]%
        {altarawy2018lascad}
\bibfield{author}{\bibinfo{person}{Doaa Altarawy}, \bibinfo{person}{Hossameldin
  Shahin}, \bibinfo{person}{Ayat Mohammed}, {and} \bibinfo{person}{Na Meng}.}
  \bibinfo{year}{2018}\natexlab{}.
\newblock \showarticletitle{Lascad : Language-agnostic software categorization
  and similar application detection}.
\newblock \bibinfo{journal}{\emph{Journal of Systems and Software}}
  \bibinfo{volume}{142} (\bibinfo{year}{2018}), \bibinfo{pages}{21--34}.
\newblock
\showISSN{0164-1212}
\urldef\tempurl%
\url{https://doi.org/10.1016/j.jss.2018.04.018}
\showDOI{\tempurl}


\bibitem[Caruana(1997)]%
        {caruana1997mtl}
\bibfield{author}{\bibinfo{person}{Rich Caruana}.}
  \bibinfo{year}{1997}\natexlab{}.
\newblock \showarticletitle{Multitask Learning}.
\newblock \bibinfo{journal}{\emph{Machine Learning}} \bibinfo{volume}{28},
  \bibinfo{number}{1} (\bibinfo{year}{1997}), \bibinfo{pages}{41--75}.
\newblock
\urldef\tempurl%
\url{https://doi.org/10.1023/A:1007379606734}
\showDOI{\tempurl}


\bibitem[Chen et~al\mbox{.}(2021)]%
        {chen-etal-2021-constructing}
\bibfield{author}{\bibinfo{person}{Catherine Chen}, \bibinfo{person}{Kevin
  Lin}, {and} \bibinfo{person}{Dan Klein}.} \bibinfo{year}{2021}\natexlab{}.
\newblock \showarticletitle{Constructing Taxonomies from Pretrained Language
  Models}. In \bibinfo{booktitle}{\emph{Proceedings of the 2021 Conference of
  the North American Chapter of the Association for Computational Linguistics:
  Human Language Technologies}}. \bibinfo{publisher}{Association for
  Computational Linguistics}, \bibinfo{address}{Online},
  \bibinfo{pages}{4687--4700}.
\newblock
\urldef\tempurl%
\url{https://doi.org/10.18653/v1/2021.naacl-main.373}
\showDOI{\tempurl}


\bibitem[Devlin et~al\mbox{.}(2019)]%
        {devlin-etal-2019-bert}
\bibfield{author}{\bibinfo{person}{Jacob Devlin}, \bibinfo{person}{Ming-Wei
  Chang}, \bibinfo{person}{Kenton Lee}, {and} \bibinfo{person}{Kristina
  Toutanova}.} \bibinfo{year}{2019}\natexlab{}.
\newblock \showarticletitle{{BERT}: Pre-training of Deep Bidirectional
  Transformers for Language Understanding}. In
  \bibinfo{booktitle}{\emph{Proceedings of the 2019 Conference of the North
  {A}merican Chapter of the ACL: HLT, Volume 1}}.
  \bibinfo{publisher}{Association for Computational Linguistics},
  \bibinfo{address}{Minneapolis, Minnesota}, \bibinfo{pages}{4171--4186}.
\newblock


\bibitem[Di~Sipio et~al\mbox{.}(2020)]%
        {sipio2020naive}
\bibfield{author}{\bibinfo{person}{Claudio Di~Sipio}, \bibinfo{person}{Riccardo
  Rubei}, \bibinfo{person}{Davide Di~Ruscio}, {and} \bibinfo{person}{Phuong~T.
  Nguyen}.} \bibinfo{year}{2020}\natexlab{}.
\newblock \showarticletitle{A Multinomial Na\"{\i}ve Bayesian ({MNB}) Network
  to Automatically Recommend Topics for GitHub Repositories}. In
  \bibinfo{booktitle}{\emph{Proceedings of the Evaluation and Assessment in
  Software Engineering}} (Trondheim, Norway) \emph{(\bibinfo{series}{EASE
  '20})}. \bibinfo{publisher}{Association for Computing Machinery},
  \bibinfo{address}{New York, NY, USA}, \bibinfo{pages}{71–80}.
\newblock
\showISBNx{9781450377317}
\urldef\tempurl%
\url{https://doi.org/10.1145/3383219.3383227}
\showDOI{\tempurl}


\bibitem[Glickman(1999)]%
        {glickman1999parameter}
\bibfield{author}{\bibinfo{person}{Mark~E Glickman}.}
  \bibinfo{year}{1999}\natexlab{}.
\newblock \showarticletitle{Parameter estimation in large dynamic paired
  comparison experiments}.
\newblock \bibinfo{journal}{\emph{Journal of the Royal Statistical Society:
  Series C (Applied Statistics)}} \bibinfo{volume}{48}, \bibinfo{number}{3}
  (\bibinfo{year}{1999}), \bibinfo{pages}{377--394}.
\newblock


\bibitem[Herbrich et~al\mbox{.}(2007)]%
        {herbrich2007trueskill}
\bibfield{author}{\bibinfo{person}{Ralf Herbrich}, \bibinfo{person}{Tom Minka},
  {and} \bibinfo{person}{Thore Graepel}.} \bibinfo{year}{2007}\natexlab{}.
\newblock \showarticletitle{TrueSkill(TM): A Bayesian Skill Rating System}. In
  \bibinfo{booktitle}{\emph{Advances in Neural Information Processing Systems
  20} (\bibinfo{edition}{advances in neural information processing systems 20}
  ed.)}. \bibinfo{publisher}{MIT Press}, \bibinfo{pages}{569--576}.
\newblock
\urldef\tempurl%
\url{https://www.microsoft.com/en-us/research/publication/trueskilltm-a-bayesian-skill-rating-system/}
\showURL{%
\tempurl}


\bibitem[Huang et~al\mbox{.}(2020)]%
        {huang2020corel}
\bibfield{author}{\bibinfo{person}{Jiaxin Huang}, \bibinfo{person}{Yiqing Xie},
  \bibinfo{person}{Yu Meng}, \bibinfo{person}{Yunyi Zhang}, {and}
  \bibinfo{person}{Jiawei Han}.} \bibinfo{year}{2020}\natexlab{}.
\newblock \showarticletitle{CoRel: Seed-Guided Topical Taxonomy Construction by
  Concept Learning and Relation Transferring}. In
  \bibinfo{booktitle}{\emph{{KDD} '20: The 26th {ACM} {SIGKDD} Conference on
  Knowledge Discovery and Data Mining, Virtual Event, CA, USA, August 23-27,
  2020}}, \bibfield{editor}{\bibinfo{person}{Rajesh Gupta},
  \bibinfo{person}{Yan Liu}, \bibinfo{person}{Jiliang Tang}, {and}
  \bibinfo{person}{B.~Aditya Prakash}} (Eds.). \bibinfo{publisher}{{ACM}},
  \bibinfo{pages}{1928--1936}.
\newblock
\urldef\tempurl%
\url{https://doi.org/10.1145/3394486.3403244}
\showDOI{\tempurl}


\bibitem[Izadi et~al\mbox{.}(2020)]%
        {izadi2020topic}
\bibfield{author}{\bibinfo{person}{Maliheh Izadi}, \bibinfo{person}{Siavash
  Ganji}, {and} \bibinfo{person}{Abbas Heydarnoori}.}
  \bibinfo{year}{2020}\natexlab{}.
\newblock \showarticletitle{Topic Recommendation for Software Repositories
  using Multi-label Classification Algorithms}.
\newblock \bibinfo{journal}{\emph{ArXiv}}  \bibinfo{volume}{abs/2010.09116}
  (\bibinfo{year}{2020}).
\newblock
\showeprint[arxiv]{2010.09116}~[cs.SE]


\bibitem[Izadi et~al\mbox{.}(2021)]%
        {izadi201repologue}
\bibfield{author}{\bibinfo{person}{Maliheh Izadi}, \bibinfo{person}{Abbas
  Heydarnoori}, {and} \bibinfo{person}{Georgios Gousios}.}
  \bibinfo{year}{2021}\natexlab{}.
\newblock \showarticletitle{Topic recommendation for software repositories
  using multi-label classification algorithms}.
\newblock \bibinfo{journal}{\emph{Empirical Software Engineering}}
  \bibinfo{volume}{26}, \bibinfo{number}{5} (\bibinfo{date}{July}
  \bibinfo{year}{2021}), \bibinfo{pages}{93}.
\newblock
\showISSN{1573-7616}
\urldef\tempurl%
\url{https://doi.org/10.1007/s10664-021-09976-2}
\showDOI{\tempurl}


\bibitem[Kawaguchi et~al\mbox{.}(2004)]%
        {Kawaguchi2006MUDABlue}
\bibfield{author}{\bibinfo{person}{Shinji Kawaguchi},
  \bibinfo{person}{Pankaj~K. Garg}, \bibinfo{person}{Makoto Matsushita}, {and}
  \bibinfo{person}{Katsuro Inoue}.} \bibinfo{year}{2004}\natexlab{}.
\newblock \showarticletitle{MUDABlue: An Automatic Categorization System for
  Open Source Repositories}. In \bibinfo{booktitle}{\emph{11th Asia-Pacific
  Software Engineering Conference {(APSEC} 2004), 30 November - 3 December
  2004, Busan, Korea}}. \bibinfo{publisher}{{IEEE} Computer Society},
  \bibinfo{pages}{184--193}.
\newblock
\urldef\tempurl%
\url{https://doi.org/10.1109/APSEC.2004.69}
\showDOI{\tempurl}


\bibitem[Krippendorff(1970)]%
        {krippendorff1970estimating}
\bibfield{author}{\bibinfo{person}{Klaus Krippendorff}.}
  \bibinfo{year}{1970}\natexlab{}.
\newblock \showarticletitle{Estimating the reliability, systematic error and
  random error of interval data}.
\newblock \bibinfo{journal}{\emph{Educational and Psychological Measurement}}
  \bibinfo{volume}{30}, \bibinfo{number}{1} (\bibinfo{year}{1970}),
  \bibinfo{pages}{61--70}.
\newblock


\bibitem[Krippendorff(2004)]%
        {krippendorff2004reliability}
\bibfield{author}{\bibinfo{person}{Klaus Krippendorff}.}
  \bibinfo{year}{2004}\natexlab{}.
\newblock \showarticletitle{Reliability in content analysis: Some common
  misconceptions and recommendations}.
\newblock \bibinfo{journal}{\emph{Human communication research}}
  \bibinfo{volume}{30}, \bibinfo{number}{3} (\bibinfo{year}{2004}),
  \bibinfo{pages}{411--433}.
\newblock


\bibitem[LeClair et~al\mbox{.}(2018)]%
        {leclair2018neural}
\bibfield{author}{\bibinfo{person}{Alexander LeClair}, \bibinfo{person}{Zachary
  Eberhart}, {and} \bibinfo{person}{Collin McMillan}.}
  \bibinfo{year}{2018}\natexlab{}.
\newblock \showarticletitle{Adapting Neural Text Classification for Improved
  Software Categorization}. In \bibinfo{booktitle}{\emph{2018 {IEEE}
  International Conference on Software Maintenance and Evolution, {ICSME} 2018,
  Madrid, Spain, September 23-29, 2018}}. \bibinfo{publisher}{{IEEE} Computer
  Society}, \bibinfo{pages}{461--472}.
\newblock
\urldef\tempurl%
\url{https://doi.org/10.1109/ICSME.2018.00056}
\showDOI{\tempurl}


\bibitem[Linares-V\'{a}squez et~al\mbox{.}(2014)]%
        {vasquez2014api}
\bibfield{author}{\bibinfo{person}{Mario Linares-V\'{a}squez},
  \bibinfo{person}{Collin Mcmillan}, \bibinfo{person}{Denys Poshyvanyk}, {and}
  \bibinfo{person}{Mark Grechanik}.} \bibinfo{year}{2014}\natexlab{}.
\newblock \showarticletitle{On Using Machine Learning to Automatically Classify
  Software Applications into Domain Categories}.
\newblock \bibinfo{journal}{\emph{Empirical Softw. Engg.}}
  \bibinfo{volume}{19}, \bibinfo{number}{3} (\bibinfo{date}{June}
  \bibinfo{year}{2014}), \bibinfo{pages}{582–618}.
\newblock
\showISSN{1382-3256}
\urldef\tempurl%
\url{https://doi.org/10.1007/s10664-012-9230-z}
\showDOI{\tempurl}


\bibitem[McMillan et~al\mbox{.}(2012)]%
        {mcmillan2012clan}
\bibfield{author}{\bibinfo{person}{Collin McMillan}, \bibinfo{person}{Mark
  Grechanik}, {and} \bibinfo{person}{Denys Poshyvanyk}.}
  \bibinfo{year}{2012}\natexlab{}.
\newblock \showarticletitle{Detecting Similar Software Applications}. In
  \bibinfo{booktitle}{\emph{Proceedings of the 34th International Conference on
  Software Engineering, {ICSE} 2012, June 2-9, 2012, Zurich, Switzerland}}
  (Zurich, Switzerland) \emph{(\bibinfo{series}{ICSE '12})}.
  \bibinfo{publisher}{{IEEE} Computer Society}, \bibinfo{pages}{364–374}.
\newblock
\showISBNx{9781467310673}
\urldef\tempurl%
\url{https://doi.org/10.1109/ICSE.2012.6227178}
\showURL{%
\tempurl}


\bibitem[Mikhailiuk et~al\mbox{.}(2020)]%
        {mikhailiuk2020active}
\bibfield{author}{\bibinfo{person}{Aliaksei Mikhailiuk},
  \bibinfo{person}{Clifford Wilmot}, \bibinfo{person}{Mar{\'{\i}}a
  P{\'{e}}rez{-}Ortiz}, \bibinfo{person}{Dingcheng Yue}, {and}
  \bibinfo{person}{Rafal~K. Mantiuk}.} \bibinfo{year}{2020}\natexlab{}.
\newblock \showarticletitle{Active Sampling for Pairwise Comparisons via
  Approximate Message Passing and Information Gain Maximization}. In
  \bibinfo{booktitle}{\emph{25th International Conference on Pattern
  Recognition, {ICPR} 2020, Virtual Event / Milan, Italy, January 10-15,
  2021}}. \bibinfo{publisher}{{IEEE}}, \bibinfo{pages}{2559--2566}.
\newblock
\urldef\tempurl%
\url{https://doi.org/10.1109/ICPR48806.2021.9412676}
\showDOI{\tempurl}


\bibitem[Moustafa et~al\mbox{.}(2018)]%
        {moustafa2018software}
\bibfield{author}{\bibinfo{person}{Sammar Moustafa}, \bibinfo{person}{Mustafa~Y
  ElNainay}, \bibinfo{person}{Nagwa El~Makky}, {and} \bibinfo{person}{Mohamed~S
  Abougabal}.} \bibinfo{year}{2018}\natexlab{}.
\newblock \showarticletitle{Software bug prediction using weighted majority
  voting techniques}.
\newblock \bibinfo{journal}{\emph{Alexandria engineering journal}}
  \bibinfo{volume}{57}, \bibinfo{number}{4} (\bibinfo{year}{2018}),
  \bibinfo{pages}{2763--2774}.
\newblock


\bibitem[Panichella et~al\mbox{.}(2013)]%
        {panichella2013ldaga}
\bibfield{author}{\bibinfo{person}{Annibale Panichella},
  \bibinfo{person}{Bogdan Dit}, \bibinfo{person}{Rocco Oliveto},
  \bibinfo{person}{Massimiliano~Di Penta}, \bibinfo{person}{Denys Poshyvanyk},
  {and} \bibinfo{person}{Andrea~De Lucia}.} \bibinfo{year}{2013}\natexlab{}.
\newblock \showarticletitle{How to effectively use topic models for software
  engineering tasks? an approach based on genetic algorithms}. In
  \bibinfo{booktitle}{\emph{35th International Conference on Software
  Engineering, {ICSE} '13, San Francisco, CA, USA, May 18-26, 2013}},
  \bibfield{editor}{\bibinfo{person}{David Notkin}, \bibinfo{person}{Betty
  H.~C. Cheng}, {and} \bibinfo{person}{Klaus Pohl}} (Eds.).
  \bibinfo{publisher}{{IEEE} Computer Society}, \bibinfo{pages}{522--531}.
\newblock
\urldef\tempurl%
\url{https://doi.org/10.1109/ICSE.2013.6606598}
\showDOI{\tempurl}


\bibitem[Runeson et~al\mbox{.}(2012)]%
        {Runeson2012threats}
\bibfield{author}{\bibinfo{person}{Per Runeson}, \bibinfo{person}{Martin
  H{\"{o}}st}, \bibinfo{person}{Austen Rainer}, {and}
  \bibinfo{person}{Bj{\"{o}}rn Regnell}.} \bibinfo{year}{2012}\natexlab{}.
\newblock \bibinfo{booktitle}{\emph{Case Study Research in Software Engineering
  - Guidelines and Examples}}.
\newblock \bibinfo{publisher}{Wiley}.
\newblock
\showISBNx{978-1-118-10435-4}
\urldef\tempurl%
\url{http://eu.wiley.com/WileyCDA/WileyTitle/productCd-1118104358.html}
\showURL{%
\tempurl}


\bibitem[Sabetta and Bezzi(2018)]%
        {sabetta2018security}
\bibfield{author}{\bibinfo{person}{Antonino Sabetta} {and}
  \bibinfo{person}{Michele Bezzi}.} \bibinfo{year}{2018}\natexlab{}.
\newblock \showarticletitle{A Practical Approach to the Automatic
  Classification of Security-Relevant Commits}. In
  \bibinfo{booktitle}{\emph{2018 {IEEE} International Conference on Software
  Maintenance and Evolution, {ICSME} 2018, Madrid, Spain, September 23-29,
  2018}}. \bibinfo{publisher}{{IEEE} Computer Society},
  \bibinfo{pages}{579--582}.
\newblock
\urldef\tempurl%
\url{https://doi.org/10.1109/ICSME.2018.00058}
\showDOI{\tempurl}


\bibitem[Sas and Capiluppi(2022)]%
        {sas2022antipatterns}
\bibfield{author}{\bibinfo{person}{Cezar Sas} {and} \bibinfo{person}{Andrea
  Capiluppi}.} \bibinfo{year}{2022}\natexlab{}.
\newblock \showarticletitle{Antipatterns in software classification
  taxonomies}.
\newblock \bibinfo{journal}{\emph{Journal of Systems and Software}}
  \bibinfo{volume}{190} (\bibinfo{year}{2022}), \bibinfo{pages}{111343}.
\newblock
\showISSN{0164-1212}
\urldef\tempurl%
\url{https://doi.org/10.1016/j.jss.2022.111343}
\showDOI{\tempurl}


\bibitem[Scarselli et~al\mbox{.}(2009)]%
        {scarselli2009gnn}
\bibfield{author}{\bibinfo{person}{Franco Scarselli}, \bibinfo{person}{Marco
  Gori}, \bibinfo{person}{Ah~Chung Tsoi}, \bibinfo{person}{Markus
  Hagenbuchner}, {and} \bibinfo{person}{Gabriele Monfardini}.}
  \bibinfo{year}{2009}\natexlab{}.
\newblock \showarticletitle{The Graph Neural Network Model}.
\newblock \bibinfo{journal}{\emph{IEEE Transactions on Neural Networks}}
  \bibinfo{volume}{20}, \bibinfo{number}{1} (\bibinfo{year}{2009}),
  \bibinfo{pages}{61--80}.
\newblock
\urldef\tempurl%
\url{https://doi.org/10.1109/TNN.2008.2005605}
\showDOI{\tempurl}


\bibitem[Shang et~al\mbox{.}(2020a)]%
        {shang-etal-2020-taxonomy}
\bibfield{author}{\bibinfo{person}{Chao Shang}, \bibinfo{person}{Sarthak Dash},
  \bibinfo{person}{Md. Faisal~Mahbub Chowdhury}, \bibinfo{person}{Nandana
  Mihindukulasooriya}, {and} \bibinfo{person}{Alfio Gliozzo}.}
  \bibinfo{year}{2020}\natexlab{a}.
\newblock \showarticletitle{Taxonomy Construction of Unseen Domains via
  Graph-based Cross-Domain Knowledge Transfer}. In
  \bibinfo{booktitle}{\emph{Proceedings of the 58th Annual Meeting of the
  Association for Computational Linguistics}}. \bibinfo{publisher}{Association
  for Computational Linguistics}, \bibinfo{address}{Online},
  \bibinfo{pages}{2198--2208}.
\newblock
\urldef\tempurl%
\url{https://doi.org/10.18653/v1/2020.acl-main.199}
\showDOI{\tempurl}


\bibitem[Shang et~al\mbox{.}(2020b)]%
        {shang2020nettaxo}
\bibfield{author}{\bibinfo{person}{Jingbo Shang}, \bibinfo{person}{Xinyang
  Zhang}, \bibinfo{person}{Liyuan Liu}, \bibinfo{person}{Sha Li}, {and}
  \bibinfo{person}{Jiawei Han}.} \bibinfo{year}{2020}\natexlab{b}.
\newblock \showarticletitle{NetTaxo: Automated Topic Taxonomy Construction from
  Text-Rich Network}. In \bibinfo{booktitle}{\emph{{WWW} '20: The Web
  Conference 2020, Taipei, Taiwan, April 20-24, 2020}},
  \bibfield{editor}{\bibinfo{person}{Yennun Huang}, \bibinfo{person}{Irwin
  King}, \bibinfo{person}{Tie{-}Yan Liu}, {and} \bibinfo{person}{Maarten van
  Steen}} (Eds.). \bibinfo{publisher}{{ACM} / {IW3C2}},
  \bibinfo{pages}{1908--1919}.
\newblock
\urldef\tempurl%
\url{https://doi.org/10.1145/3366423.3380259}
\showDOI{\tempurl}


\bibitem[Sharma et~al\mbox{.}(2017)]%
        {sharma2017cataloging}
\bibfield{author}{\bibinfo{person}{Abhishek Sharma}, \bibinfo{person}{Ferdian
  Thung}, \bibinfo{person}{Pavneet~Singh Kochhar}, \bibinfo{person}{Agus
  Sulistya}, {and} \bibinfo{person}{David Lo}.}
  \bibinfo{year}{2017}\natexlab{}.
\newblock \showarticletitle{Cataloging GitHub Repositories}. In
  \bibinfo{booktitle}{\emph{Proceedings of the 21st International Conference on
  Evaluation and Assessment in Software Engineering}} (Karlskrona, Sweden)
  \emph{(\bibinfo{series}{EASE'17})}. \bibinfo{publisher}{Association for
  Computing Machinery}, \bibinfo{address}{New York, NY, USA},
  \bibinfo{pages}{314–319}.
\newblock
\showISBNx{9781450348041}
\urldef\tempurl%
\url{https://doi.org/10.1145/3084226.3084287}
\showDOI{\tempurl}


\bibitem[Soll and Vosgerau(2017)]%
        {soll2017classifyhub}
\bibfield{author}{\bibinfo{person}{Marcus Soll} {and} \bibinfo{person}{Malte
  Vosgerau}.} \bibinfo{year}{2017}\natexlab{}.
\newblock \showarticletitle{ClassifyHub: An Algorithm to Classify GitHub
  Repositories}. In \bibinfo{booktitle}{\emph{KI 2017: Advances in Artificial
  Intelligence}}, \bibfield{editor}{\bibinfo{person}{Gabriele Kern-Isberner},
  \bibinfo{person}{Johannes F{\"u}rnkranz}, {and} \bibinfo{person}{Matthias
  Thimm}} (Eds.). \bibinfo{publisher}{Springer International Publishing},
  \bibinfo{address}{Cham}, \bibinfo{pages}{373--379}.
\newblock
\showISBNx{978-3-319-67190-1}


\bibitem[Tian et~al\mbox{.}(2009)]%
        {tian2009lact}
\bibfield{author}{\bibinfo{person}{Kai Tian}, \bibinfo{person}{Meghan Revelle},
  {and} \bibinfo{person}{Denys Poshyvanyk}.} \bibinfo{year}{2009}\natexlab{}.
\newblock \showarticletitle{Using Latent Dirichlet Allocation for automatic
  categorization of software}. In \bibinfo{booktitle}{\emph{Proceedings of the
  6th International Working Conference on Mining Software Repositories, {MSR}
  2009 (Co-located with ICSE), Vancouver, BC, Canada, May 16-17, 2009,
  Proceedings}}, \bibfield{editor}{\bibinfo{person}{Michael~W. Godfrey} {and}
  \bibinfo{person}{Jim Whitehead}} (Eds.). \bibinfo{publisher}{{IEEE} Computer
  Society}, \bibinfo{pages}{163--166}.
\newblock
\urldef\tempurl%
\url{https://doi.org/10.1109/MSR.2009.5069496}
\showDOI{\tempurl}


\bibitem[Ugurel et~al\mbox{.}(2002)]%
        {Ugurel2002classification}
\bibfield{author}{\bibinfo{person}{Secil Ugurel}, \bibinfo{person}{Robert
  Krovetz}, {and} \bibinfo{person}{C.~Lee Giles}.}
  \bibinfo{year}{2002}\natexlab{}.
\newblock \showarticletitle{What's the Code? Automatic Classification of Source
  Code Archives}. In \bibinfo{booktitle}{\emph{Proceedings of the Eighth ACM
  SIGKDD International Conference on Knowledge Discovery and Data Mining}}
  (Edmonton, Alberta, Canada) \emph{(\bibinfo{series}{KDD '02})}.
  \bibinfo{publisher}{Association for Computing Machinery},
  \bibinfo{address}{New York, NY, USA}, \bibinfo{pages}{632–638}.
\newblock
\showISBNx{158113567X}
\urldef\tempurl%
\url{https://doi.org/10.1145/775047.775141}
\showDOI{\tempurl}


\bibitem[Vargas{-}Baldrich et~al\mbox{.}(2015)]%
        {vargas2015automatic}
\bibfield{author}{\bibinfo{person}{Santiago Vargas{-}Baldrich},
  \bibinfo{person}{Mario~Linares V{\'{a}}squez}, {and} \bibinfo{person}{Denys
  Poshyvanyk}.} \bibinfo{year}{2015}\natexlab{}.
\newblock \showarticletitle{Automated Tagging of Software Projects Using
  Bytecode and Dependencies {(N)}}. In \bibinfo{booktitle}{\emph{30th
  {IEEE/ACM} International Conference on Automated Software Engineering, {ASE}
  2015, Lincoln, NE, USA, November 9-13, 2015}},
  \bibfield{editor}{\bibinfo{person}{Myra~B. Cohen}, \bibinfo{person}{Lars
  Grunske}, {and} \bibinfo{person}{Michael Whalen}} (Eds.).
  \bibinfo{publisher}{{IEEE} Computer Society}, \bibinfo{pages}{289--294}.
\newblock
\urldef\tempurl%
\url{https://doi.org/10.1109/ASE.2015.38}
\showDOI{\tempurl}


\bibitem[V{\'{a}}squez et~al\mbox{.}(2016)]%
        {linares2016clandroid}
\bibfield{author}{\bibinfo{person}{Mario~Linares V{\'{a}}squez},
  \bibinfo{person}{Andrew Holtzhauer}, {and} \bibinfo{person}{Denys
  Poshyvanyk}.} \bibinfo{year}{2016}\natexlab{}.
\newblock \showarticletitle{On automatically detecting similar Android apps}.
  In \bibinfo{booktitle}{\emph{24th {IEEE} International Conference on Program
  Comprehension, {ICPC} 2016, Austin, TX, USA, May 16-17, 2016}}.
  \bibinfo{publisher}{{IEEE} Computer Society}, \bibinfo{pages}{1--10}.
\newblock
\urldef\tempurl%
\url{https://doi.org/10.1109/ICPC.2016.7503721}
\showDOI{\tempurl}


\bibitem[Vrande\v{c}i\'{c}(2012)]%
        {vrandecic2012wikidata}
\bibfield{author}{\bibinfo{person}{Denny Vrande\v{c}i\'{c}}.}
  \bibinfo{year}{2012}\natexlab{}.
\newblock \showarticletitle{Wikidata: A New Platform for Collaborative Data
  Collection}. In \bibinfo{booktitle}{\emph{Proceedings of the 21st
  International Conference on World Wide Web}} (Lyon, France)
  \emph{(\bibinfo{series}{WWW '12 Companion})}. \bibinfo{publisher}{Association
  for Computing Machinery}, \bibinfo{address}{New York, NY, USA},
  \bibinfo{pages}{1063–1064}.
\newblock
\showISBNx{9781450312301}
\urldef\tempurl%
\url{https://doi.org/10.1145/2187980.2188242}
\showDOI{\tempurl}


\bibitem[Wang et~al\mbox{.}(2017)]%
        {wang-etal-2017-short}
\bibfield{author}{\bibinfo{person}{Chengyu Wang}, \bibinfo{person}{Xiaofeng
  He}, {and} \bibinfo{person}{Aoying Zhou}.} \bibinfo{year}{2017}\natexlab{}.
\newblock \showarticletitle{A Short Survey on Taxonomy Learning from Text
  Corpora: Issues, Resources and Recent Advances}. In
  \bibinfo{booktitle}{\emph{Proceedings of the 2017 Conference on Empirical
  Methods in Natural Language Processing}}. \bibinfo{publisher}{Association for
  Computational Linguistics}, \bibinfo{address}{Copenhagen, Denmark},
  \bibinfo{pages}{1190--1203}.
\newblock
\urldef\tempurl%
\url{https://doi.org/10.18653/v1/D17-1123}
\showDOI{\tempurl}


\bibitem[Ye and Doermann(2014)]%
        {ye2014subjective}
\bibfield{author}{\bibinfo{person}{Peng Ye} {and} \bibinfo{person}{David~S.
  Doermann}.} \bibinfo{year}{2014}\natexlab{}.
\newblock \showarticletitle{Active Sampling for Subjective Image Quality
  Assessment}. In \bibinfo{booktitle}{\emph{2014 {IEEE} Conference on Computer
  Vision and Pattern Recognition, {CVPR} 2014, Columbus, OH, USA, June 23-28,
  2014}}. \bibinfo{publisher}{{IEEE} Computer Society},
  \bibinfo{pages}{4249--4256}.
\newblock
\urldef\tempurl%
\url{https://doi.org/10.1109/CVPR.2014.541}
\showDOI{\tempurl}


\bibitem[Yu et~al\mbox{.}(2020)]%
        {yu-etal-2020-hearst}
\bibfield{author}{\bibinfo{person}{Changlong Yu}, \bibinfo{person}{Jialong
  Han}, \bibinfo{person}{Peifeng Wang}, \bibinfo{person}{Yangqiu Song},
  \bibinfo{person}{Hongming Zhang}, \bibinfo{person}{Wilfred Ng}, {and}
  \bibinfo{person}{Shuming Shi}.} \bibinfo{year}{2020}\natexlab{}.
\newblock \showarticletitle{When Hearst Is not Enough: Improving Hypernymy
  Detection from Corpus with Distributional Models}. In
  \bibinfo{booktitle}{\emph{Proceedings of the 2020 Conference on Empirical
  Methods in Natural Language Processing (EMNLP)}}.
  \bibinfo{publisher}{Association for Computational Linguistics},
  \bibinfo{address}{Online}, \bibinfo{pages}{6208--6217}.
\newblock
\urldef\tempurl%
\url{https://doi.org/10.18653/v1/2020.emnlp-main.502}
\showDOI{\tempurl}


\bibitem[Zhang et~al\mbox{.}(2018)]%
        {zhang2018taxogen}
\bibfield{author}{\bibinfo{person}{Chao Zhang}, \bibinfo{person}{Fangbo Tao},
  \bibinfo{person}{Xiusi Chen}, \bibinfo{person}{Jiaming Shen},
  \bibinfo{person}{Meng Jiang}, \bibinfo{person}{Brian~M. Sadler},
  \bibinfo{person}{Michelle Vanni}, {and} \bibinfo{person}{Jiawei Han}.}
  \bibinfo{year}{2018}\natexlab{}.
\newblock \showarticletitle{TaxoGen: Unsupervised Topic Taxonomy Construction
  by Adaptive Term Embedding and Clustering}. In
  \bibinfo{booktitle}{\emph{Proceedings of the 24th {ACM} {SIGKDD}
  International Conference on Knowledge Discovery {\&} Data Mining, {KDD} 2018,
  London, UK, August 19-23, 2018}}, \bibfield{editor}{\bibinfo{person}{Yike
  Guo} {and} \bibinfo{person}{Faisal Farooq}} (Eds.).
  \bibinfo{publisher}{{ACM}}, \bibinfo{pages}{2701--2709}.
\newblock
\urldef\tempurl%
\url{https://doi.org/10.1145/3219819.3220064}
\showDOI{\tempurl}


\bibitem[Zhang et~al\mbox{.}(2019)]%
        {zhang2019HiGitClass}
\bibfield{author}{\bibinfo{person}{Yu Zhang}, \bibinfo{person}{Frank~F. Xu},
  \bibinfo{person}{Sha Li}, \bibinfo{person}{Yu Meng}, \bibinfo{person}{Xuan
  Wang}, \bibinfo{person}{Qi Li}, {and} \bibinfo{person}{Jiawei Han}.}
  \bibinfo{year}{2019}\natexlab{}.
\newblock \showarticletitle{HiGitClass: Keyword-Driven Hierarchical
  Classification of GitHub Repositories}. In \bibinfo{booktitle}{\emph{2019
  {IEEE} International Conference on Data Mining, {ICDM} 2019, Beijing, China,
  November 8-11, 2019}}, \bibfield{editor}{\bibinfo{person}{Jianyong Wang},
  \bibinfo{person}{Kyuseok Shim}, {and} \bibinfo{person}{Xindong Wu}} (Eds.).
  \bibinfo{publisher}{{IEEE}}, \bibinfo{pages}{876--885}.
\newblock
\urldef\tempurl%
\url{https://doi.org/10.1109/ICDM.2019.00098}
\showDOI{\tempurl}


\bibitem[Zhou et~al\mbox{.}(2015)]%
        {zhou2015clstm}
\bibfield{author}{\bibinfo{person}{Chunting Zhou}, \bibinfo{person}{Chonglin
  Sun}, \bibinfo{person}{Zhiyuan Liu}, {and} \bibinfo{person}{Francis C.~M.
  Lau}.} \bibinfo{year}{2015}\natexlab{}.
\newblock \showarticletitle{A {C-LSTM} Neural Network for Text Classification}.
\newblock \bibinfo{journal}{\emph{CoRR}}  \bibinfo{volume}{abs/1511.08630}
  (\bibinfo{year}{2015}).
\newblock
\showeprint[arxiv]{1511.08630}
\urldef\tempurl%
\url{http://arxiv.org/abs/1511.08630}
\showURL{%
\tempurl}


\bibitem[Zhou et~al\mbox{.}(2021)]%
        {9590294}
\bibfield{author}{\bibinfo{person}{Yuqi Zhou}, \bibinfo{person}{Jiawei Wu},
  {and} \bibinfo{person}{Yanchun Sun}.} \bibinfo{year}{2021}\natexlab{}.
\newblock \showarticletitle{GHTRec: {A} Personalized Service to Recommend
  GitHub Trending Repositories for Developers}. In
  \bibinfo{booktitle}{\emph{2021 {IEEE} International Conference on Web
  Services, {ICWS} 2021, Chicago, IL, USA, September 5-10, 2021}},
  \bibfield{editor}{\bibinfo{person}{Carl~K. Chang}, \bibinfo{person}{Ernesto
  Daminai}, \bibinfo{person}{Jing Fan}, \bibinfo{person}{Parisa Ghodous},
  \bibinfo{person}{Michael Maximilien}, \bibinfo{person}{Zhongjie Wang},
  \bibinfo{person}{Robert Ward}, {and} \bibinfo{person}{Jia Zhang}} (Eds.).
  \bibinfo{publisher}{{IEEE}}, \bibinfo{pages}{314--323}.
\newblock
\urldef\tempurl%
\url{https://doi.org/10.1109/ICWS53863.2021.00049}
\showDOI{\tempurl}


\end{thebibliography}

\end{document}